\documentclass[aps, pra, groupedaddress, superscriptaddress, amsmath, amssymb, twocolumn, longbibliography]{revtex4-2}
\usepackage{amsmath}

\usepackage{qcircuit}
\usepackage{braket}  
\usepackage{color}
\usepackage{booktabs}
\usepackage{multirow}
\usepackage{tikz}
\usepackage{lipsum}
\usepackage{orcidlink}
\usepackage{adjustbox}
\usepackage{hyperref}
\usepackage{subfigure}

\begin{document}

\title{Improved performance of the Bacon-Shor code with Steane's syndrome extraction method}

\author{Guillermo Escobar-Arrieta \orcidlink{0009-0009-7061-6824}}
\email{guillermo.escobar@ucr.ac.cr}
\affiliation{Escuela de Física, Universidad de Costa Rica, San José 2060, Costa Rica}
\affiliation{Escuela de Ingenier\'ia El\'ectrica, Universidad de Costa Rica, San José 2060, Costa Rica}

\author{Mauricio Gutiérrez \orcidlink{0000-0002-0826-3521}}
\email{mauricio.gutierrez\_a@ucr.ac.cr}
\affiliation{Escuela de Química, Universidad de Costa Rica, San José 2060, Costa Rica}

\begin{abstract}
We compare Steane's and Shor's syndrome extraction methods on the Bacon-Shor code.  We propose a straightforward strategy based on post-selection to prepare the logical $|0\rangle_L$ and $|+\rangle_L$ states of the Bacon-Shor code by using flag-like qubits to verify their constituent Greenberger-Horne-Zeilinger states.  We perform stabilizer simulations with a circuit-level, depolarizing Pauli error model and find that Steane's method significantly outperforms Shor's.  Not only does Steane's method result in pseudo-thresholds that are about 1 order of magnitude higher than Shor's, but also its advantage increases monotonically as we go from a distance-3 to a distance-9 Bacon-Shor code.  The advantage of Steane's method is the greatest in the regime where gate errors dominate over measurement errors.  Some of the circuit constructions we propose for Steane's method are not formally fault-tolerant, yet outperform the formally fault-tolerant Shor's protocols for experimentally relevant physical error rates.  This suggests that constructing formally fault-tolerant circuits that maintain the full code distance is not strictly necessary to guarantee the usefulness of a quantum error-correcting protocol.  Despite relying on post-selection, we find that our methods can be efficient.  These protocols would be naturally implementable on a platform with long-range qubit interactions like trapped ions or neutral atoms.     
\end{abstract}

\maketitle

\section{Introduction}

Quantum error correction (QEC) \cite{ShorCode, TheoryQECC, Shor96_1, Calderbank97, KitaevAnyons, TerhalReviewQEC} will be a crucial component to construct a large-scale, fault-tolerant (FT) quantum computer capable of solving certain problems that are prohibitively costly on classical computers \cite{ShorFactoring, Lloyd96, Alan2005}.  In order to build logical qubits with an error rate sufficiently low to allow for the implementation of deep quantum circuits, quantum error-correcting codes (QECC) will be employed.  QECCs encode logical qubits in a larger number of physical qubits.  If the error rate affecting the physical qubits is below the threshold of the particular QECC being employed, then an arbitrarily low logical error rate can be obtained with only a polylogarithmic overhead in physical resources \cite{Aharonov96, KnillThreshold, PreskillThreshold, Burkard2005, Nielsen2005, Leung2006, AliferisThreshold}.  

In stabilizer codes, the logical codespace is defined as the simultaneous ($+1$)-eigenspace of a set of independent, commuting Pauli operators known as the stabilizer generators \cite{Calderbank97, Calderbank98, GottesmanThesis}.  Errors are detected by extracting the error syndrome, \textit{i.e.}, the eigenvalues of the stabilizer generators.  After decoding the error syndrome, a conditional correction is applied on the logical qubits.

Since errors can occur on every qubit and after every gate, the circuit constructions used to extract the error syndrome need to be FT.  There are several methods to achieve this.  Shor's syndrome extraction method (ShorSEM) employs ancillary qubits to measure the stabilizer generators one by one.  To guarantee fault tolerance, faults should not propagate from the ancillary qubits to the data qubits to create uncorrectable errors.  This can be achieved in a variety of ways.  In the original proposal \cite{Shor96_1, Shor96_2}, each stabilizer generator would be measured with a Greenberger-Horne-Zeilinger (GHZ) state, either pre-verified or subsequently decoded to correct correlated errors arising during its preparation \cite{AliferisSlowMeas, Surfacecodetwist}.  More recently, it has been shown that a single ancilla qubit can be FT if it is coupled to extra flag qubits used to identify the harmful errors that have propagated to the data qubits \cite{ChaoFlagPRL, ChaoFlagPRX}.  Furthermore, for certain stabilizer codes, single ancillary qubits without flag qubits are sufficient to measure each stabilizer generator fault-tolerantly \cite{TomitaSvore, StanleyAlonzo, LiDirectBS}.  Finally, the stabilizer generators need to be measured several times to guarantee that readout errors or data errors that occur between stabilizer measurements do not become fatal.  Notable improvements on this considerable time overhead have been designed since Shor's original proposal \cite{Shor96_1}, including adaptive \cite{Zalka, DelfosseShorEC, InkAdaptive} and single-shot \cite{BombinSingleShot, CampbellSingleShot} methods. 

For Calderbank-Shor-Steane (CSS) codes, an alternative scheme to ShorSEM is Steane's syndrome extraction method (SteaneSEM) \cite{SteaneEC}.  In this scheme, an ancillary logical $|0 \rangle_L$ ($|+\rangle_L$) state of the same code as the data block's code is used to extract the error syndrome associated with the X (Z) stabilizers by coupling it to the data logical qubit by means of a logical CNOT, performing a logical measurement in the X (Z) basis and finally classically error-correcting the outcome.  As long as the preparation of the ancillary logical states is FT, it is sufficient to perform this procedure only once since the logical CNOT gate is transversal on CSS codes.  SteaneSEM reduces the time overhead of ShorSEM at the expense of the requirement of FT preparation of the logical $|0\rangle_L$ and $|+\rangle_L$ states. 
 In fact, rather than disparate schemes, these two methods can be regarded as opposite ends of a family of circuit constructions that exchange the complexity of the ancilla block for a reduction in the number of repetitions necessary to guarantee fault tolerance \cite{ShilinShorSteanePRA, ShilinShorSteanePRL}.  Recently, SteaneSEM has been experimentally demonstrated in two different trapped-ion systems \cite{DukeSteaneEC, InnsbruckSteaneEC}.  In a similar spirit to SteaneSEM, yet applicable to any stabilizer code, Knill's syndrome extraction method (KnillSEM) employs a logical Bell pair to extract all the error syndrome in one step \cite{KnillEC}.              
 
 Two-dimensional Bacon-Shor (BS) codes \cite{BSOriginal, ShorCode, BSAliferisCross} are a family of CSS codes defined on a planar array.  The logical subspace has dimension higher than 2 and thus contains several logical qubits, which can be seen as subsystems of the codespace.  Out of these subsystems, typically only one is chosen as the working logical qubit and the rest are referred to as gauge (logical) qubits.  Crucially for this work, depending on the state of these gauge qubits, the logical $|0\rangle_L$ and $|+\rangle_L$ states of a BS code can be expressed as products of GHZ states.  This vastly simplifies the FT preparation of the logical $|0\rangle_L$ and $|+\rangle_L$ states and thus positions the BS code as a natural candidate for SteaneSEM.  
 
 Here we study how to fault-tolerantly prepare logical $|0\rangle_L$ and $|+\rangle_L$ states of the BS code by using extra qubits to verify their constituent GHZ states and post-selecting them.  We go up to distance-9.  For BS codes of distances 3 to 9, we compare the logical error rate for 1 QEC cycle with ShorSEM and SteaneSEM.  For ShorSEM, we employ single ancillary qubits to directly measure the high-weight stabilizers \cite{LiDirectBS} and use a recently developed adaptive protocol \cite{InkAdaptive} for the time decoding.   For the space decoding of both methods, we use a lookup table.  Although exponentially costly in the limit of large code distance, the lookup-table decoder for the BS code is essentially the same as for a repetition code and, therefore, does not scale as badly as for subspace codes.  

In some papers in the literature, ShorSEM exclusively refers to the case where the ancillary qubits are prepared in a GHZ (cat) state, which guarantees fault tolerance for a general stabilizer code.  Since for the BS code it is possible to guarantee fault tolerance with a single ancillary physical qubit for each stabilizer generator, throughout this paper ShorSEM will refer to the case where a single ancillary qubit (and not a cat state) is used to measure each stabilizer generator.

We find that SteaneSEM outperforms ShorSEM for all distances greater than 3.  More importantly, the advantage of SteaneSEM over ShorSEM grows monotonically with the distance and the pseudo-thresholds remain about 1 order of magnitude higher for the former than the latter.  We also find that the advantage of SteaneSEM over ShorSEM is the highest when gate errors dominate over measurement errors.  Finally, we calculate the probability of the GHZ states not passing the verification and find it to be very reasonable.  The fact that each GHZ state can be separately employed in SteaneSEM without the need to simultaneously have all the constituent GHZ states of the logical state makes this scheme very efficient despite its post-selective nature.  

This paper is organized as follows.  In Sections \ref{sec:Steane} and \ref{sec:BScode}, we review SteaneSEM and the BS code, respectively.  Section \ref{sec:SimulationScheme} contains the relevant details of our simulation scheme, including the noise model and the importance sampler we employ.  In Section \ref{sec:Results}, we summarize our main results.  Finally, in Section \ref{sec:Conclusions}, we conclude and present some open questions and future directions.






\section{Steane's Syndrome Extraction Method (SteaneSEM)} \label{sec:Steane}

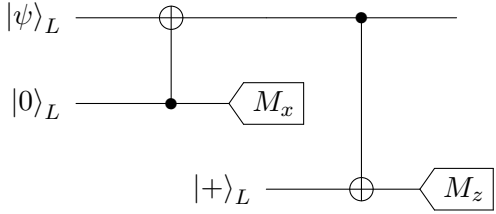
\begin{figure}[h]

    \[\Qcircuit @C=0.7em @R=.4em @! {
    \lstick{\ket{\psi}_L} & \targ & \qw & \ctrl{2} & \qw \\
    \lstick{\ket{0}_L} & \ctrl{-1} & \measuretab{M_{x}} & & \\
    & & \lstick{\ket{+}_L} & \targ & \measuretab{M_{z}}  \\
    }\]
        
    \caption{Circuit representation of SteaneSEM used to correct for errors on a logical qubit in an arbitrary state $|\psi\rangle_L$.  Each CNOT in the circuit corresponds to a logical CNOT composed of $n$ physical CNOTs for a CSS code with $n$ data qubits.  The ancillary logical qubit prepared in $|0\rangle_L$ ($|+\rangle_L$) is used to correct Z(X) errors which propagate down from the data logical qubit through the CNOTs.  After coupling this ancillary logical qubit to the data logical qubit, the former is measured in the X(Z) basis and the outcome is processed with classical error correction. It is not necessary to repeat the procedure in order to achieve fault tolerance as long as the ancillary logical states are prepared in a FT way.}
    \label{SteaneEC}
    \vspace{-5pt}
\end{figure}

SteaneSEM is a single-shot method to extract the true syndrome of a CSS code.  The procedure is depicted in Figure \ref{SteaneEC}.

The main challenge of SteaneSEM lies in the FT preparation of the logical $|0\rangle_L$ and $|+\rangle_L$ states that are needed to correct for Z and X errors, respectively.  In general, preparing a logical $|0\rangle_L$ ($|+\rangle_L$) state of a CSS code requires a considerable overhead.  One can initialize the qubits in $|0\rangle^{\otimes n}$ ($|+\rangle^{\otimes n}$), where $n$ is the total number of data qubits, and then measure the X(Z) stabilizer generators to project the product state to the codespace.  However, to make this procedure FT, the stabilizer measurements need to be repeated, which essentially amounts to performing ShorSEM.  More efficient methods have been developed, which employ classical error-correcting codes to aid in the detection of problematic errors \cite{Brun2017, Brun2018}.  However, as we show in the next section, for the BS code, the FT logical state preparation can be even more straightforward.

When performing SteaneSEM, there are two mechanisms by which errors can propagate from the ancillary to the data qubits: either directly or indirectly.  For the ancillary $|0\rangle_L$ ($|+\rangle_L$) state, X(Z) errors can directly propagate through the logical CNOT to the data qubits.  On the other hand, while Z(X) errors will not propagate directly through the logical CNOT, they will affect the measurements and will propagate indirectly by giving rise to an incorrect syndrome and the application of the wrong correction.  Therefore, it is of critical importance to prepare the ancillary logical states in a FT fashion.

\section{The Bacon-Shor (BS) code}
\label{sec:BScode}

\begin{figure*}
    \centering
    \subfigure[d=3]{%
        \includegraphics[width=0.25\textwidth]{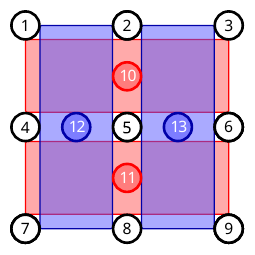}
    }
    \hfill
    \subfigure[stabilizers]{\raisebox{2cm}{%
        \includegraphics[width=0.10\textwidth]{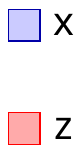}
    }}
    \hfill
    \subfigure[d=5]{%
        \includegraphics[width=0.45\textwidth]{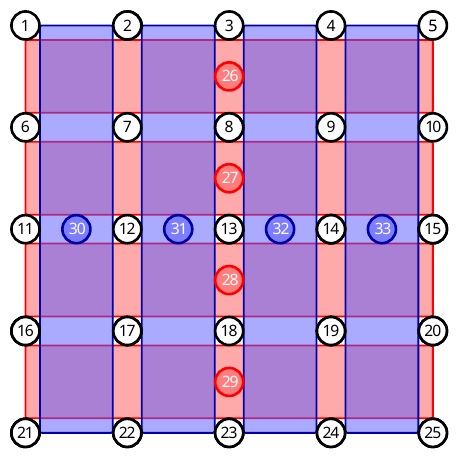}
    }
    \caption{Square BS codes of distance (a) 3 and (c) 5, respectively. Each code contains $d^2$ data qubits (black circles), as well as $(d-1)$ X stabilizer generators (blue vertical rectangles) and $(d-1)$ Z stabilizer generators (red horizontal rectangles).}
    \label{BSCodeFigure}
\end{figure*}

BS codes \cite{BSOriginal, ShorCode, BSAliferisCross} are a family of CSS codes defined on a planar array of physical qubits.  BS codes are subsystem codes \cite{OQEC1, OQEC2}. 
This implies that the logical subspace has dimension higher than 2 and thus contains several logical qubits, which can be seen as subsystems of the codespace.  Out of these subsystems, typically only one is chosen as the working logical qubit.

For symmetric or square BS codes, there are $d^2$ physical data qubits, where $d$ is the length of the side of the lattice and the distance of the code.  There are $(d-1)$ X stabilizer generators and $(d-1)$ Z stabilizer generators, each of weight $2d$.  X(Z) stabilizer generators correspond to vertical (horizontal) rectangles.  Because there are $d^2$ physical data qubits but only $2(d-1)$ stabilizer generators, the total number of logical qubits is $d^2 - 2(d-1) > 1$.  One of these has distance $d_L = d$ and is used as the actual logical qubit.  The rest correspond to encoded qubits of distance $d_G < d$ that are not used to store information.  They are referred to as gauge qubits because they can be regarded as gauge degrees of freedom.  The BS code is an instance of the more general 2-D compass code family \cite{CompassCodes}.  Figure \ref{BSCodeFigure} illustrates the square BS codes of distances $d=5$ and $d=3$, with their respective stabilizer generators.


The BS code does not present a quantum error correction threshold as the lattice size is increased \cite{BSNoThreshold}.  However, if the code distance is increased by concatenation, a threshold is obtained, albeit a small one \cite{BSAliferisCross}.  More importantly, under certain experimental conditions, simulations of the BS code have revealed that it can achieve a comparable and even superior performance to the more popular surface code \cite{Dripto2020}.  Furthermore, it has been recently shown that by employing a construction based on lattice surgery a threshold can be obtained for the BS code \cite{LessBacon}.

The BS code has several very useful properties.  Every stabilizer can be measured fault-tolerantly with a single bare ancilla as long as one is careful about the ordering of the entangling gates \cite{LiDirectBS}.  This property is very useful for systems which allow for long-range entangling gates, like trapped ions \cite{Egan2021} and neutral atoms \cite{DolevNeutral}.  Alternatively, the stabilizers can also be measured indirectly by measuring their constituent gauge operators (or some combination of them) and calculating their total parity \cite{BSAliferisCross}, which makes the BS code amenable to be implemented on solid-state systems with only nearest-neighboring interactions. 

All logical Pauli gates, CNOT$_L$, $H_L$, and $Y(\pi/2)_L$ can be implemented transversally on the BS code.  Universality can be achieved with either a $T_L$ gate via magic state distillation \cite{MagicStateOriginal} or a $CCZ_L$ gate by means of pieceable fault tolerance \cite{YoderBSCCZ, PieceableFT}. The BS code has been shown to be useful against leakage errors \cite{LeakageNBrown}, amenable to be run with continuous measurements \cite{BSContinuous}, and capable of hosting extra dynamical logical qubits by an appropriate scheduling of the check operators \cite{BSFloquet}.  The distance-2 and distance-3 square BS codes have been experimentally implemented in trapped-ion systems \cite{422, Egan2021}.   

The gauge degrees of freedom of the BS code can be exploited to design very simple FT procedures to prepare the logical $|0\rangle_L$ and $|+\rangle_L$ states.  In particular, for a square [[$d^2, 1, d$]] BS code, the logical states can be expressed as: 

\begin{align*}
 |0\rangle_L &= \frac{1}{\sqrt{2^d}} \left( |+\rangle^{\otimes d} + |-\rangle^{\otimes d} \right) ^{\otimes d} \quad \mbox{ along rows} \\
  |+\rangle_L  &= \frac{1}{\sqrt{2^d}} \left( |0\rangle^{\otimes d} + |1\rangle^{\otimes d} \right) ^{\otimes d} \quad \mbox{ along columns}
\end{align*}

\begin{figure}[h]
    \[ 
    \Qcircuit @C=0.7em @R=.4em @! {
    \lstick{\ket{0}} & \gate{H} & \ctrl{1} & \qw & \qw & \qw & \qw\\
    \lstick{\ket{0}} & \qw & \targ & \ctrl{1} & \qw & \qw  & \qw\\
    \lstick{\ket{0}} & \qw & \qw & \targ & \ctrl{1} & \qw  & \qw\\
    \lstick{\ket{0}} & \qw & \qw & \qw & \targ & \ctrl{1} & \qw\\
    \lstick{\ket{0}} & \qw & \qw & \qw & \qw & \targ & \qw\\
    }\]
        
    \caption{Circuit that can be used to prepare the GHZ state $\frac{1}{\sqrt{2}} \left( |0\rangle^{\otimes 5} + |1\rangle^{\otimes 5} \right)$.  For the noise model employed in this paper, this circuit is not FT, since a weight-1 X error can propagate through the CNOTs to form a weight-2 X error.  Z errors are not problematic since they propagate through the CNOTs to form an operator which is actually a stabilizer of the GHZ state.}
    \label{GHZ}
    \vspace{-5pt}
\end{figure}
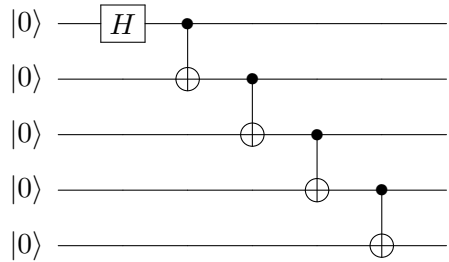

That is, to prepare the logical $|0\rangle_L$ $\left( |+\rangle_L \right)$ state, we only need to prepare the GHZ state $\left( |+\rangle^{\otimes d} + |-\rangle^{\otimes d} \right)/\sqrt{2}$ $\left( \left( |0\rangle^{\otimes d} + |1\rangle^{\otimes d} \right) / \sqrt{2} \right)$ on each one of the $d$ rows (columns) of the code.  The preparation of GHZ states for QEC has been experimentally demonstrated in trapped-ion systems \cite{NorbertGHZion}.    Figure \ref{GHZ} illustrates a possible circuit for the preparation of $\left( |0\rangle^{\otimes 5} + |1\rangle^{\otimes 5} \right) / \sqrt{2}$.  

\subsection{FT preparation of logical $|0\rangle_L$ and $|+\rangle_L$ states on the [[25,1,5]] BS code (d=5)}


    

\begin{figure}
    \includegraphics[width=0.5\textwidth]{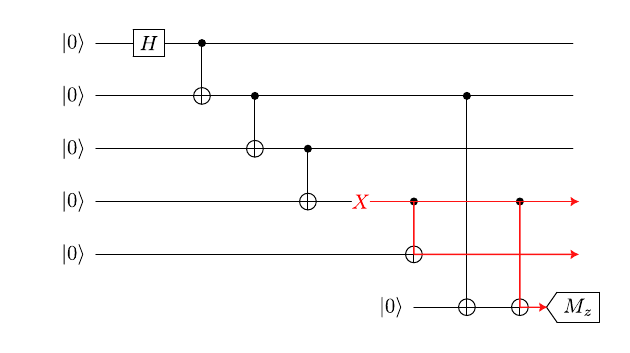}
    \centering
    \caption{The preparation of a 5-qubit GHZ state can be made FT by adding an extra verification qubit, which measures one of the Z stabilizers of the GHZ state.  In this case, the measured stabilizer is $Z_2Z_4$, but this choice is not unique.  Any stabilizer that detects a weight-2 X error caused by a single-qubit X error works well.  For example, for this circuit, $Z_1Z_5$ would also be a useful verification stabilizer.  As with any verification protocol, the final state might be discarded even in the absence of problematic errors, for instance if an error occurs on the measurement.  In the case depicted, a problematic X error is successfully detected and the GHZ is discarded and re-prepared.}
    \label{VerifiedGHZ}
\end{figure}

The construction of a 3-qubit GHZ state of the form $\frac{1}{\sqrt{2}} \left( |0\rangle^{\otimes 3} + |1\rangle^{\otimes 3} \right)$ is naturally FT because its stabilizer generators are $\{ Z_1Z_2, Z_2Z_3, X_1X_2X_3 \}$, which means that X and Z errors can be at most of weight-1 up to a stabilizer.  On the other hand, the analogous 5-qubit GHZ state, whose preparation is depicted in Figure \ref{GHZ} is stabilized by $\{ Z_1Z_2$, $Z_2Z_3$, $Z_3Z_4$, $Z_4Z_5$, $X_1X_2X_3X_4X_5 \}$.  All Z errors are at most of weight-1 up to a stabilizer, so they are not problematic but X errors can be of weight-1 or weight-2.  These are problematic because a single X error on the control qubit of a CNOT can propagate to form a weight-2 X error. 

As shown in Figure \ref{VerifiedGHZ}, this can be easily detected with an extra verification qubit.  If each one of the 5 GHZ states is verified by an extra qubit, then the resulting BS logical state will be FT and amenable to be used as an ancilla in SteaneSEM.  The procedure is straightforward and the only caveat is that it requires post-selection and, therefore, midcircuit measurements.  If the verification is not passed, the GHZ state needs to be re-prepared.  


\subsection{FT preparation of logical $|0\rangle_L$ and $|+\rangle_L$ states on the [[49,1,7]] BS code (d=7)}

The preparation of 3-qubit GHZ states for the purpose of SteaneSEM on the [[9,1,3]] BS code does not require a verification because, up to the stabilizers of the GHZ state, all weight-1 errors propagate to form other weight-1 errors.  As shown in the previous subsection, for 5-qubit GHZ states, some weight-1 errors can propagate to form problematic weight-2 errors.  Fortunately, these can be caught with a single verification qubit.

\begin{figure}
    \includegraphics[width=0.5\textwidth]{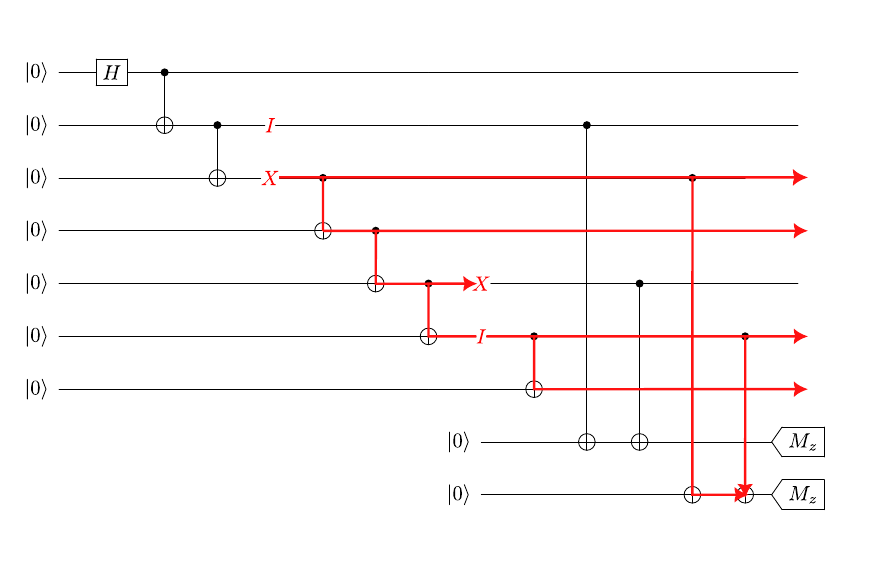}
    \centering
    \caption{Circuit to verify a 7-qubit GHZ with 2 verification qubits.  For the error model we employ, this circuit is formally not FT.  In this case, an X error that occurs with probability $O(p^2)$ can cause an undetected weight-3 X error (The resulting weight-4 error is equivalent to a weight-3 error up to the GHZ's X stabilizer.).  As shown in the Section \ref{sec:Results}, even though this circuit is not formally FT, it still performs better than ShorSEM for $p > 2 \times 10^{-6}$.  To guarantee formal fault tolerance, we find that 3 verification qubits are needed.}
    \label{GHZd7v2} 
\end{figure}

For the [[49,1,7]] code, in order to maintain the distance-7, the situation becomes more complex: one needs to make sure that (1) no weight-1 error propagates to form a problematic weight-2 or weight-3 error, and also that (2) no weight-2 error propagates to form a problematic weight-3 error.  Naively, by extrapolating from the first two cases, one might believe that 2 verification qubits are sufficient.  However, this is not the case.  We performed an exhaustive search over all the possible circuits that emply 2 verification qubits to measure weight-2 Z stabilizers and found none that satisfies conditions (1) and (2).  Figure \ref{GHZd7v2} depicts an instance of a circuit with 2 verification qubits that does not satisfy the condition (2).  The conditions are not satisfied because our error model assumes that weight-2 errors can occur after entangling gates with the same probability $p$ as weight-1 errors.  If weight-2 errors after entangling gates occurred with a probability $O(p^2)$, then 2 verification qubits would suffice.  Therefore, when adapting these ideas to experimental systems, it is crucial to consider the specific noise model because the number of necessary verification qubits might depend on it.

For the error model employed in this paper, 3 verification qubits are necessary to guarantee the FT preparation of the 7-qubit GHZ state.  All 3 verification measurements need to return a $+1$ eigenvalue to accept the GHZ state.  If at least 1 of them returns a $-1$, the verification is not passed and the GHZ state needs to be re-prepared.  This was found by an exhaustive search.  Similarly, it was proven that the conditions (1) and (2) are satisfied by exhaustively checking that all weight-1 and weight-2 errors (more accurately all errors that occur with probability $O(p)$ and $O(p^2)$) that result in problematic higher-weight errors are effectively caught by at least one verification qubit.

\subsection{Maintaining the formal distance of the code is not strictly necessary to suppress the logical error rate}

When constructing QEC circuits it is usually assumed that to guarantee fault tolerance, one needs to maintain the code distance.  What is the point of using a distance-7 QEC code if a particular circuit construction is not immune to some weight-3 errors, like the case of the [[49,1,7]] code with 7-qubit GHZ states with 2 verification qubits?  It seems like a loss because the leading order of the logical error rate would go from $p^4$ to $p^3$.  

However, this argument is only valid in the limit of $p \to 0$.  For higher $p$ values, the coefficients of the non-leading-order terms might play a very significant role.  In fact, as shown in Section \ref{sec:Results}, for the [[49,1,7]] BS code with SteaneSEM, employing 2 verification qubits outperforms ShorSEM for an experimentally relevant interval of $p$ values ($p > 2 \times 10^{-6}$) despite not being formally FT.  Furthermore, for $p > 4 \times 10^{-5}$, employing 2 and 3 verification qubits result in similar performances (See Figure \ref{alld5}).  Intuitively, this occurs because, with 2 verification qubits, there are very few weight-3 errors that result in a logical error, so the coefficient of the $p^3$ term is much smaller than the coefficient of the $p^4$ term for the procedure with ShorSEM.  The coefficients of the leading orders of the polynomial expansions of the logical error rates are presented in Table \ref{tab:SteaneLeadingOrders}.  These coefficients are not obtained by curve fitting, but rather by employing an error subset sampler described in the Section \ref{sec:SimulationScheme}.  They are exact up to the sampling error of the subsets.

\subsection{FT preparation of logical $|0\rangle_L$ and $|+\rangle_L$ states on the [[81,1,9]] BS code (d=9) and beyond}

A $d$-qubit GHZ state ($(|0 \rangle^{\otimes d} + |1 \rangle^{\otimes d})/\sqrt{2}$) can have at most $1$ Z error and $\lfloor (d-1)/2 \rfloor$ X errors, up to its stabilizers.  Therefore, for errors of weight up to $\lfloor (d-1)/2 \rfloor$, no errors of higher weight will be formed by gate propagation when creating the $d$-qubit GHZ states.  We exhaustively searched over all possible constructions involving $3$ and $4$ verification qubits and we did not find a circuit that maintains the formal distance-9.  However, the performance is good compared to ShorSEM, as shown in Section \ref{sec:Results}.  All the circuits that we used to verify the GHZ states are presented in Appendix \ref{app:Circs}.  These circuits were found by an exhaustive search of all circuits that measure weight-2 Z stabilizers of the GHZ states.

\section{Simulation scheme}
\label{sec:SimulationScheme}

We use a simulation toolkit \cite{GithubRepo}, similar to the one employed in previous papers \cite{StanleyAlonzo, FaultPathTracer, TroutSurfaceIon, TransversalCNOT}, with CHP \cite{CHP} as its core simulator.  In order to compare SteaneSEM and ShorSEM on the BS code, we simulate 1 QEC cycle.  In both cases, we initialize the data qubits in a perfect logical state, perform one of the syndrome extraction methods, and apply the corresponding correction to the data qubits.  Finally, to account for only uncorrectable errors, we project the corrected state back to the codespace.  We count a logical error if the final projected state is different from the initial state.  We also determine if the working logical qubit ends up being entangled with the gauge logical qubits.  For the noise model we employ, logical entanglement never occurs.  We performed the simulations with an initial logical $|0\rangle_L$ and $|+\rangle_L$.  We show the results for $|+\rangle_L$, but the results are practically the same for both initial states.  


For SteaneSEM, to calculate the logical error rate, we take into account only the runs where all the GHZ verifications were passed.  To perform the classical error correction on the ancillary logical state outcomes, we employ a lookup table.  Despite its exponential scaling with the number of stabilizer generators, the lookup table is very practical given the reduced number of stabilizer generators of the BS code.  For example, for the largest code we analyze ($d=9$), the number of syndromes in each lookup table is only $2^{9-1} = 256$.  The lookup tables for the distance-$3$ and distance-$5$ BS codes are presented in Tables \ref{ltd3} and \ref{ltd5} in the Appendix.  Generalizing the lookup tables to higher distances is straightforward, since they are equivalent to the lookup tables of the repetition code.    

\begin{figure}[h]
    \centering
    \includegraphics[width=0.5\textwidth]{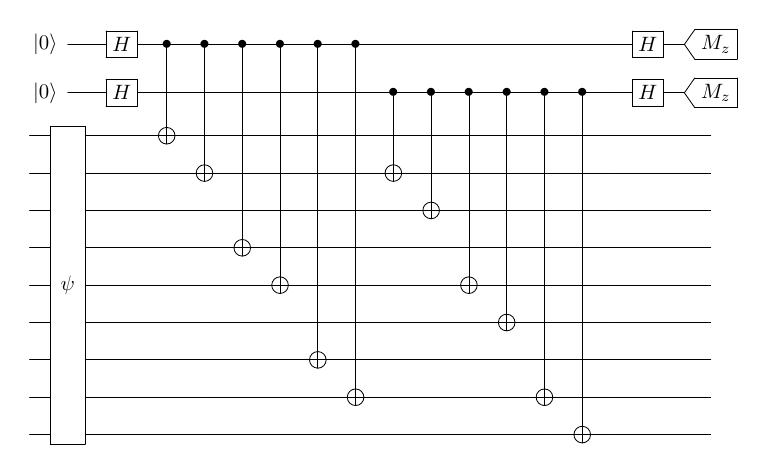}
    \caption{Circuit used to extract the X-stabilizer syndromes using ShorSEM for a distance-3 Bacon-Shor code whose state is $|\psi \rangle$.  The first two qubits are the ancillary qubits used to measure the stabilizer generators $X_1X_2X_4X_5X_7X_8$ and $X_2X_3X_5X_6X_8X_9$, respectively.  These two generators are depicted in Figure \ref{BSCodeFigure}.  A single bare qubit is sufficient to measure the generators in a FT way as long as the order of the entangling gates prevents harmful hook errors \cite{LiDirectBS}.  We use an analogous circuit to extract the Z-stabilizer syndromes.  To guarantee fault tolerance, these circuits are then repeated several times depending on the previous measurement outcomes.  Since the time decoding protocols are adaptive, the number of rounds of stabilizer measurements is not fixed from the start.  For both the weak and strong ShorSEM, the minimal number of rounds is proportional to $d$, while the maximal number of rounds is proportional to $d^2$.  The exact values are presented in Table \ref{tab:repsShorSEM}.}
    \label{fig:shorSEMd3}
\end{figure}

For ShorSEM, we employ a recently proposed adaptive scheme \cite{InkAdaptive} for the time decoding and the lookup tables for the space decoding.  Figure \ref{fig:shorSEMd3} depicts a representative ShorSEM subcircuit used in the simulations.  By keeping track in real time of the differences of syndrome measurement outcomes from consecutive rounds, the adaptive time decoder can estimate the minimal number of errors that occurred and decide when to stop.  This is more efficient than the original ShorSEM, which requires, in the worst case, $(t+1)^2$ rounds of full syndrome measurements, where $t \leq \lfloor (d-1)/2 \rfloor$ and $d$ is the code distance  The adaptive time decoder considerably reduces the total number of rounds of stabilizer measurements, although the worst case is still proportional to $t^2$.  The exact number of rounds for best and worst cases are presented in Table \ref{tab:repsShorSEM}.  The adaptive time decoder is applicable to any stabilizer code and can be employed to construct either a strong FT or a weak FT error correction protocol.

In general, an error correction protocol is FT if it satisfies two conditions, the error correction correctness property (ECCP) and the error correction recovery property (ECRP).  The ECCP refers to the ability of the protocol to maintain the correctability of an encoded state even in the presence of faults during its execution.  The ECRP refers to the property of the protocol to not increase the effective weight of the error on the input state by more than the number of faults that occurred during its execution even if the output state is still correctable.  Strong and weak fault tolerance are defined by two different degrees of stringency of the ECRP.  The precise definitions of the ECCP and the ECRP for both strong and weak fault tolerance are presented in Table \ref{tab:FTdefinitions}.  We refer to \cite{InkAdaptive} for a more detailed explanation.  

\begin{table*}[]
\centering
\noindent\makebox[\textwidth]{\begin{tabular}{|c| p{7cm} | p{7cm} |}
\hline
                                & ECCP & ECRP \\ \hline
    strongly $t$-fault tolerant & For any input encoded state with an error of weight $r$, if $s$ faults occur during the protocol's execution (with $r + s \leq t$), then ideally decoding the input and output states give the same encoded state.   &  For any input encoded state, if $s$ faults occur during the protocol's execution (with $s \leq t$), regardless of the weight of the error on the input state, the output state differs from any valid encoded state by an error of weight at most $s$.    \\ \hline
    weakly $t$-fault tolerant   & Same as above.     &  For any input encoded state with an error of weight $r$, if $s$ faults occur during the protocol's execution (with $r+s \leq t$), the output state differs from any valid encoded state by an error of weight at most $s$.    \\ \hline  
\end{tabular}}
\caption{The error correction correctness property (ECCP) and the error correction recovery property (ECRP) that define the notions of strongly and weakly FT ShorSEM used in this paper.  The distinction between strong and weak fault tolerance lies on different degrees of stringency of the ECRP.  Strong fault tolerance is necessary for concatenation.  For a more detailed discussion, we refer the reader to \cite{InkAdaptive}.}
\label{tab:FTdefinitions}
\end{table*}

\subsection{Noise model}

We employ a depolarizing Pauli noise model with no memory errors.  Specifically, we have the following noise processes:
\begin{enumerate}
    \item After every 1-qubit unitary gate: X, Y, or Z error, each with a probability of $p/3$.
    \item After every 2-qubit unitary gate: one of the 15 possible Pauli errors (IX, IY, ..., ZZ), each with a probability of $p/15$.
    \item After every $|0\rangle$ state preparation: an X error with a probability of $p$.
    \item After every measurement in the Z basis: a bit flip with a probability of $q$.  For most of the discussion, we set $p=q$ to simplify the visualization of the results.  At the end, we analyze the case where $p$ and $q$ are independent to explore how each syndrome extraction method performs under different gate vs. measurement noise strengths.  
\end{enumerate}

Despite being restricted to only Clifford gates and Pauli preparations and measurements, error channels within the stabilizer formalism can be good approximations to realistic non-Clifford noise processes \cite{Cory1, Mau1, Cory2, Mau2, Mau3}.

\subsection{Importance sampler}

To expedite the simulation, we use an importance (subset) sampler previously employed \cite{StanleyAlonzo, TransversalCNOT}.  Instead of traversing the whole circuit and adding an error after each gate if a randomly generated real number between 0 and 1 is less than the physical error probability $p$, our importance sampler divides the error-configuration set into non-overlapping subsets according to the number of errors or weight.  

For a noise model with $n$ independent error parameters, we label each subset with a vector $\vec{w} = (w_1, w_2, ..., w_n)$, where $w_i$ corresponds to the number of errors associated with the parameter $i$.  To estimate the logical error rate, we (1) analytically calculate the total probability of occurrence of each subset ($A_{\vec{w}}$) and (2) perform Monte Carlo sampling on the error subsets with high probability of occurrence to obtain the logical error rate for each subset $\left( p_L^{(\vec{w})} \right)$.  We can then compute lower and upper bounds to the logical error rate with the following equations:
\begin{equation}
    p_{L}^{(lower)} = \sum_{\vec{w} = (0,0,...,0)}^{\vec{w}_{max}} A_{\vec{w}} \, p_L^{(\vec{w})}
    \label{eq:pl}
\end{equation}
\begin{equation}
    p_{L}^{(upper)} = p_{L}^{(lower)} + \left(1 - \sum_{\vec{w} = (0,0,...,0)}^{\vec{w}_{max}} A_{\vec{w}} \right)
\end{equation}
where $\vec{w}_{max}$ is the highest-weight subset that was sampled.  The lower (upper) bound assumes that all the subsets not sampled have a logical error rate of 0(1).  To calculate the upper bound we simply add to the lower bound the total probability of occurrence of the unsampled subsets.  For low physical error rates, the two bounds typically overlap.  They start to diverge as the physical error rates increase.  

The subset sampler has several very convenient features.  First, it allows us to efficiently obtain logical error rates at extremely low physical error rates.  In fact, it is most efficient at low physical error rates because the number of subsets needed to be sampled is low.  Secondly, once we sample the relevant subsets, we can then generate the whole logical error rate curve (or hyper-surface for a multi-parameter noise model) at once by simply re-calculating the probabilities of occurrence of each subset, which is done analytically.  Finally, it allows us to compute the coefficients associated with each term in the polynomial expansion of the logical error rate, which is very useful to compare different QEC schemes.  The method to compute the polynomial coefficients is described in the Appendix \ref{app:Coeff} and the coefficients of the leading order terms for all logical error rates are reported in Table \ref{tab:SteaneLeadingOrders}.  We can even obtain the exact values of the leading polynomial coefficients by exhaustively running all the error configurations of the relevant subsets, as long as their cardinalities are not prohibitively high.  Other importance samplers have been previously applied to the simulation of QEC circuits \cite{AlexRareEvents, GuttentagDataSyndrome}.  Recently, these ideas have been extended to develop a dynamical subset sampling scheme \cite{SaschaDynamicalSubset}.  

In this work, we assume 3 independent error weights: the number of errors after 1-qubit gates and state preparations ($w_1$), after 2-qubit gates ($w_2$), and after measurements ($w_3$).  However, to simplify the visualization of the results, we set all the physical error rates to be equal, except in the final part of the paper where we let the measurement error rate to be independent.   We sample all error subsets up to total weight $w_{tot} = w_1 + w_2 + w_3$ equal to 10.  The number of samples taken for each subset is $2 \times 10^4$, except for crucial subsets with very a low logical error rate, where we take $\max(2 \times 10^4, 5 \%$ of the subset's cardinality$)$.

\section{Results} \label{sec:Results}

\begin{table*}[]
\centering
\noindent\makebox[\textwidth]{\begin{tabular}{|l|ll|lllll|}
\hline
\multirow{2}{*}{d} & \multicolumn{2}{c|}{Shor}                                                 & \multicolumn{5}{c|}{Steane}                                                                                                                                                                                                                                                                                                                                                                                     \\ \cline{2-8} 
                   & \multicolumn{1}{c|}{Weak}                   & \multicolumn{1}{c|}{Strong} & \multicolumn{1}{l|}{$v=0$}                   & \multicolumn{1}{l|}{$v=1$}                   & \multicolumn{1}{l|}{$v=2$}                                                                                     & \multicolumn{1}{l|}{$v=3$}                                                                                      & $v=4$                                                                                    \\ \hline
3                  & \multicolumn{1}{l|}{$1.45\times10^{2}p^2$}  & $3.88\times10^{2}p^2$       & \multicolumn{1}{l|}{$3.66\times10^{2}p^2$} & \multicolumn{1}{l|}{--}                    & \multicolumn{1}{l|}{--}                                                                                      & \multicolumn{1}{l|}{--}                                                                                       & --                                                                                     \\ \hline
5                  & \multicolumn{1}{l|}{$2.65\times10^{5}p^3$}  & $2.13\times10^{5}p^3$       & \multicolumn{1}{l|}{$1.31\times10^{5}p^2$} & \multicolumn{1}{l|}{$4.56\times10^{4}p^3$} & \multicolumn{1}{l|}{--}                                                                                      & \multicolumn{1}{l|}{--}                                                                                       & --                                                                                     \\ \hline
7                  & \multicolumn{1}{l|}{$7.36\times10^{8}p^4$}  & $3.42\times10^{8}p^4$       & \multicolumn{1}{l|}{--}                    & \multicolumn{1}{l|}{--}                    & \multicolumn{1}{l|}{\begin{tabular}[c]{@{}l@{}}$1.31\times10^{3}p^3$\\ +\\ $6.32\times10^{7}p^4$\end{tabular}} & \multicolumn{1}{l|}{$9.88\times10^{7}p^4$}                                                                    & --                                                                                      \\ \hline
9                  & \multicolumn{1}{l|}{$1.45\times10^{11}p^5$} & $1.97\times10^{12}p^5$      & \multicolumn{1}{l|}{--}                    & \multicolumn{1}{l|}{--}                    & \multicolumn{1}{l|}{--}                                                                                      & \multicolumn{1}{l|}{\begin{tabular}[c]{@{}l@{}}$1.78\times10^{5}p^4$\\ +\\ $3.85\times10^{10}p^5$\end{tabular}} & \begin{tabular}[c]{@{}l@{}}$2.26\times10^4p^4$\\ +\\ $4.56\times10^{10}p^5$\end{tabular} \\ \hline
\end{tabular}}
\caption{Leading-order terms of the polynomial expansions of the logical error rates for BS codes of distances 3, 5, 7, and 9 and various syndrome extraction methods.  For SteaneSEM, the number of qubits used for the GHZ verifications is denoted by $v$. For each term, the coefficient is obtained with the subset sampler and the method described in Appendix \ref{app:Coeff}.  For both the strong and weak StrongSEM, the leading orders are $p^{(d+1)/2}$, which implies that the full code distance is maintained.  For SteaneSEM, the full code distance is maintained only for $d=5 (v=1)$ and $d=7 (v=3)$.  For the other cases, the leading order is reduced by 1.  Nevertheless, as shown in Figure \ref{alld5}, except for the $d=5$ without verification, the SteaneSEM cases that have a decreased distance still outperform full-distance ShorSEM for experimentally relevant physical error rates $\left( p > \sim 10^{-6} \right)$.  This illustrates that maintaining the full distance is not strictly necessary to guarantee that a particular protocol will be effective.}
\label{tab:SteaneLeadingOrders}
\end{table*}

\begin{figure*}[hbt!]
\centering
\includegraphics[width=\textwidth]{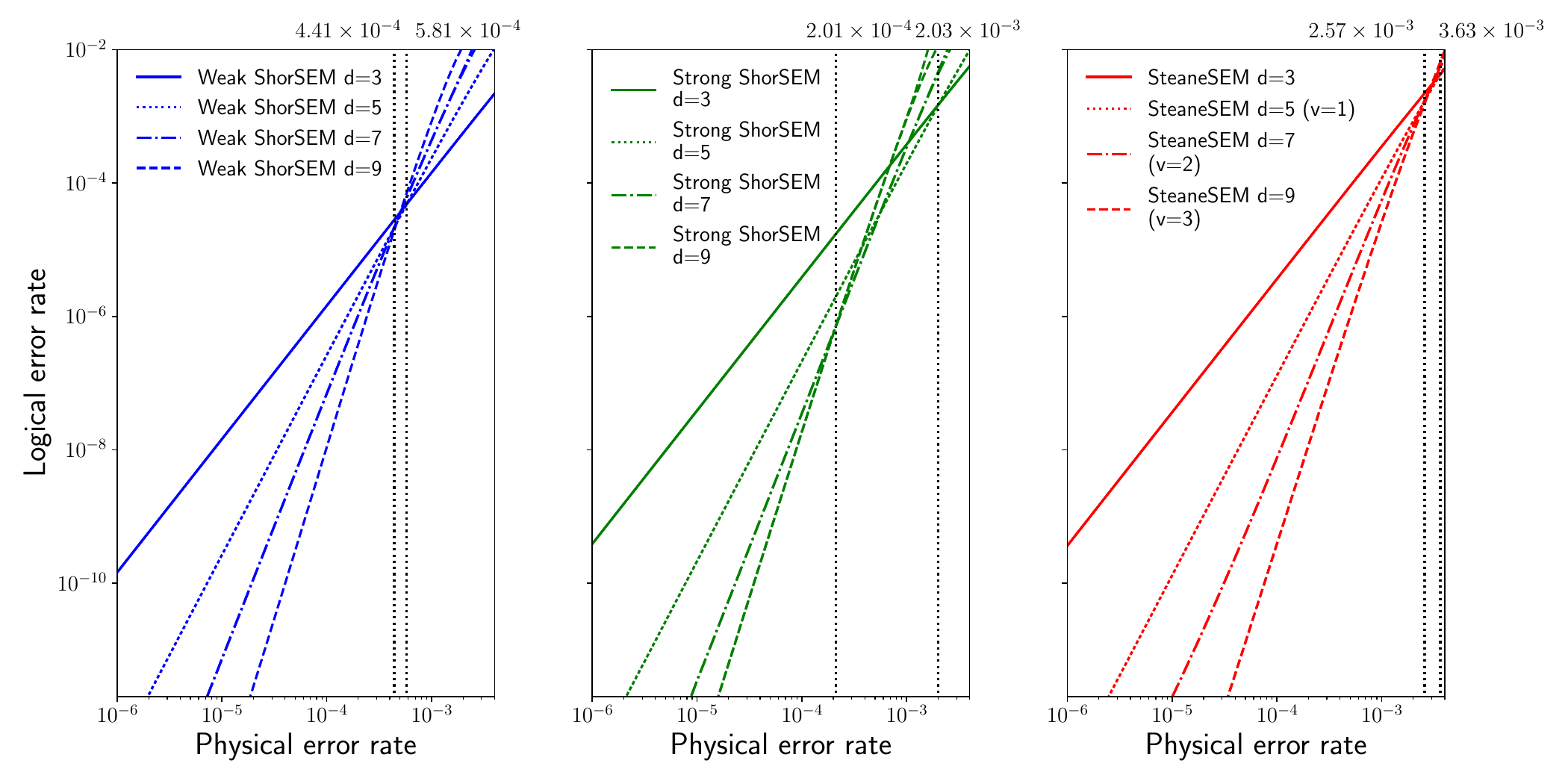}
\caption{Logical error rates for the BS code with 3 different decoding strategies: (1) weak ShorSEM adaptive decoder, (2) strong ShorSEM adaptive decoder, and (3) SteaneSEM. For each decoding strategy, the lowest and highest pseudo-thresholds are shown.  Remarkably, the lowest SteaneSEM pseudo-thresholds $\left( \sim 10^{-3} \right)$ are about 1 order of magnitude higher than their ShorSEM counterparts $\left( \sim 10^{-4} \right)$.  For SteaneSEM, the number of verification qubits used for each GHZ state is denoted by $v$.  For some $d=9$ curves, we already observe, for high values of $p$, a small divergence between the upper and lower bounds of the logical error rate.  This occurs because, for larger physical error rates, the probability of occurrence of the high-weight subsets not sampled becomes considerable.}
\label{mainplot}
\end{figure*}

Figure \ref{mainplot} shows the logical error rate for several distances of the BS code with 3 different decoding strategies.   For both the weak and strong ShorSEM, the leading order of the polynomial expansion for each curve is $p^{(d+1)/2}$ (See Table \ref{tab:SteaneLeadingOrders}), which implies that the full distance is maintained.  However, the pseudo-thresholds (intersections between curves) are rather low.  An interesting difference is observed between strong and weak ShorSEM.  Whereas the pseudo-thresholds decrease quickly for the strong decoder (from $2.0 \times 10^{-3}$ to $2.0 \times 10^{-4}$), they decrease much more slowly for the weak decoder (from $5.8 \times 10^{-4}$ to $4.4 \times 10^{-4}$).  

On the other hand, for SteaneSEM, the pseudo-thresholds remain high (from $3.6 \times 10^{-3}$ to $2.6 \times 10^{-3}$), about 1 order of magnitude higher than for the Shor methods.  It remains an open question how fast the SteaneSEM pseudo-thresholds will decrease for higher distances ($d>9$).  In any case, despite its great importance from a theoretical QEC perspective, from a practical point of view the existence of a threshold is not strictly necessary since the crucial goal of a QECC is to achieve a sufficiently low logical error rate useful for algorithmic purposes.     

\begin{figure*}[t]
  \vspace{0pt}
    \centering
    \includegraphics[width= \textwidth]{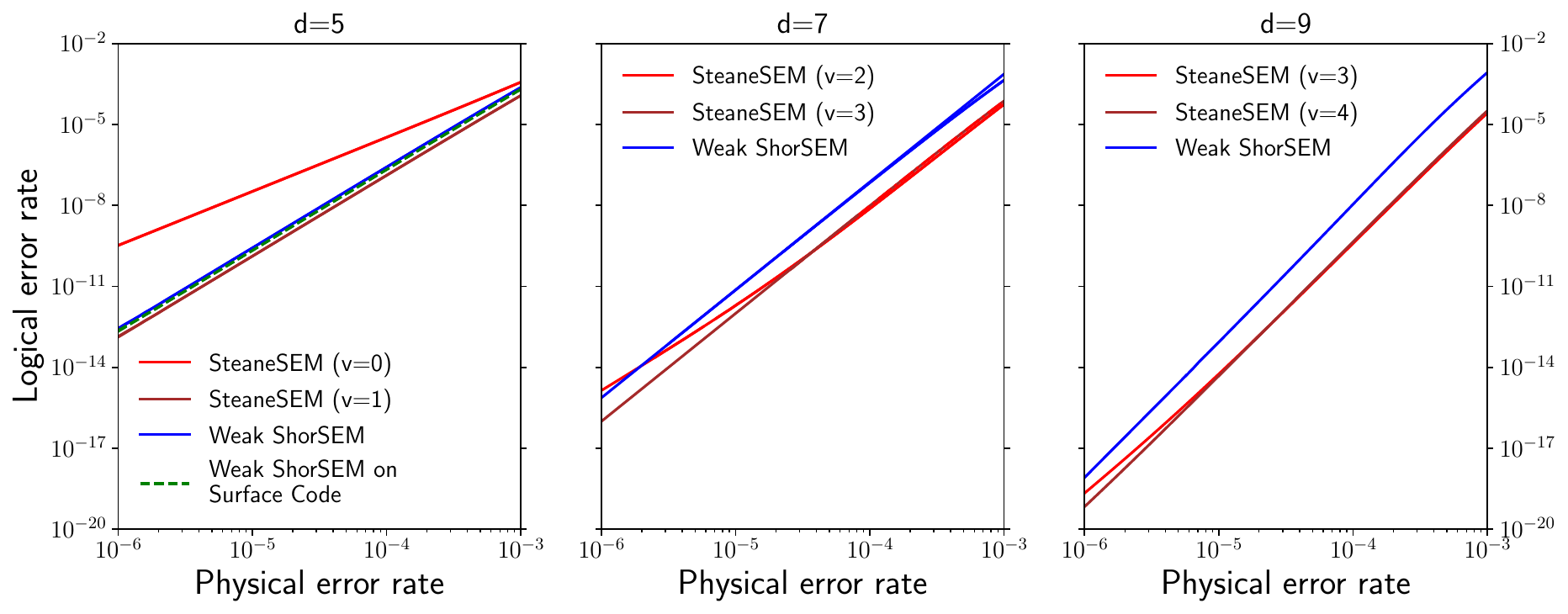}       
    \vspace{0pt}
    \caption{Logical error rates for several distances (d) of the BS code.  For SteaneSEM, the number of verification qubits used for each GHZ state is denoted by $v$.  Except for the distance-5 with no verification qubits, all the non FT SteaneSEM circuit constructions outperform the FT ShorSEM protocols for experimentally relevant values of $p$.   We only include the results for weak ShorSEM.  The results for strong ShorSEM are very similar.}
    \label{alld5}
    \vspace{0pt}
\end{figure*}


Figure \ref{alld5} presents the logical error rates from a different perspective, which lets us compare more easily the performance of the syndrome extraction methods for distances 5, 7, and 9.  For $d=5$, SteaneSEM with 1 verification qubit per GHZ state ($v=1$) outperforms ShorSEM for every physical error rate.  For low $p$ values, this can be explained by comparing the coefficients of the leading orders in the polynomial expansions (See Table \ref{tab:SteaneLeadingOrders}).  For both ShorSEM and SteaneSEM ($v=1$), the leading order is $p^{3}$, but the coefficient is almost 1 order of magnitude larger for both Shor methods than for SteaneSEM.   

As a reference point, we also simulated the $d=5$ rotated surface code \cite{TomitaSvore, Dennis2002, BombinComparative, Cleland2012} with the adaptive ShorSEM time decoder and a lookup-table space decoder.  As seen in Figure \ref{alld5}, its performance is slightly better than the BS code with ShorSEM, but still worse than the BS code with SteaneSEM. 

For $d=7$, SteaneSEM requires 3 verification qubits per GHZ state ($v=3$) to maintain the full distance of the code.  If only 2 verification qubits are used, then some weight-3 error events can be uncorrectable.  An example of such an event would be a situation where first the 2 X errors depicted in Figure \ref{GHZd7v2} occur during the preparation of one of the GHZ states.  The resulting weight-3 X error on the GHZ state error would not directly propagate to the logical data qubit because this GHZ state is a constituent of the logical $|+\rangle_L$ state (See Figure \ref{SteaneEC}).  However, it would give rise to an incorrect syndrome, so it propagates indirectly.  If another X error occurs on one of the data qubits not part of this wrong correction, then applying the correction would result in an uncorrectable weight-4 X error.  Therefore, this is an error event that occurs with probability $O(p^3)$, but that results in an uncorrectable weight-4 error.

Despite being formally not FT (its effective distance decreased from 7 to 5), SteaneSEM with 2 verification qubits per GHZ still outperforms both strong and weak ShorSEM for an experimentally relevant physical error rate interval $\left( p > 2 \times 10^{-6} \right)$, as shown in Figure \ref{alld5}.  As seen in Table \ref{tab:SteaneLeadingOrders}, the coefficient of the $p^3$ term for $d=7$ ($v=2$) SteaneSEM is very small compared to the coefficients of the $p^4$ terms for both strong and weak ShorSEM ($\sim 10^3$ vs. $\sim 10^8$).  In other words, for $d=7$ ($v=2$) SteaneSEM, the number of uncorrectable error events that occur with probability $O(p^3)$ is so small that it is only for very low physical error rates $\left( p < 2 \times 10^{-6} \right)$ that the leading-order comparison is an appropriate analysis tool.

For $d=7$, if we use 3 verification qubits per GHZ state, then SteaneSEM outperforms ShorSEM for every physical error rate.  This agrees with the fact that both $d=7$ ($v=3$) SteaneSEM and $d=7$ ShorSEM have the same leading order ($p^4$), but the former has a lower coefficient (See Table \ref{tab:SteaneLeadingOrders}).  Surprisingly, for $p > 4 \times 10^{-5}$, the performance of SteaneSEM is about the same for $v=2$ and $v=3$.  This further strengthens the point that maintaining the full distance of the code is not strictly necessary to guarantee the usefulness of a particular QEC protocol.  

For $d=9$, both $v=3$ and $v=4$ SteaneSEM have their effective distances decrease from 9 to 7, as seen from the leading $p^4$ orders in Table \ref{tab:SteaneLeadingOrders}.  However, as seen in Figure \ref{alld5}, for $p > 4 \times 10^{-7}$, both SteaneSEM constructions outperform their ShorSEM counterparts, despite the latter ones maintaining the full code distance ($d=9$).  Similarly to the $d=7$ case, for an experimentally relevant physical error rate interval $\left( p > 10^{-5} \right)$, using 3 or 4 verification qubits per GHZ gives essentially the same results.  

As Figure \ref{alld5} shows, the advantage of the formally FT protocols starts to become noticeable for extremely low physical error rates ($p < 10^{-6}$).  It is highly unlikely that experimental quantum systems will ever achieve such extremely low physical error rates for operations like entangling gates and measurements.  For experimentally realistic physical error rates ($p > 10^{-5}$), the FT protocols offer no advantage over their non-FT counterparts.  Therefore, to scale up our method to larger code distances, we do not envision searching for formally FT verification circuits, which would be very demanding by brute-force search.  Rather, we envision fixing the number of verification qubits to a constant value (for instance 3 or 4) and searching for the verification circuits that result in the lowest logical error rate.  Furthermore, depending on the noise model, it might not be necessary to perform very complex verifications.  For example, in several systems dephasing is the leading source of errors \cite{NiggScience,LucasIons,RosenblumScience,TaylorNature}.  In this case, it would not even be necessary to have very stringent verifications for GHZ states in the Z basis, since any dephasing error would amount to a 1-qubit Z error.

It is important to note that we are not the first to observe a good performance of non-distance-preserving QEC protocols under experimentally relevant physical error rates.  Gidney has found that, by compiling the surface code with pair measurements, its performance improves for $p=10^{-3}$ and its threshold increases, despite its distance decreasing \cite{GidneyPentagons}.  More recently, Gidney and Jones have proposed two different circuit constructions for the color code that fail to preserve the full code distance, but result in superior performances for $p=10^{-3}$ and the color code's highest threshold to date \cite{GidneyColorCodes}.  Other authors have also found good performances for the color code with non-distance-preserving QEC protocols \cite{CostUniversality, ChamberlandTriangularColorCodes, FacilitatingFTColorCodes}.  Distance-preserving protocols \cite{BalintDistancePreserving} will likely play a role for very low physical error rates where the leading order of the logical error rate's polynomial expansion becomes overwhelmingly strong.



\subsection{Improvement rate of SteaneSEM  vs. ShorSEM}

\begin{figure}[t]
  \vspace{0pt}
    \centering
    \includegraphics[width=0.45 \textwidth]{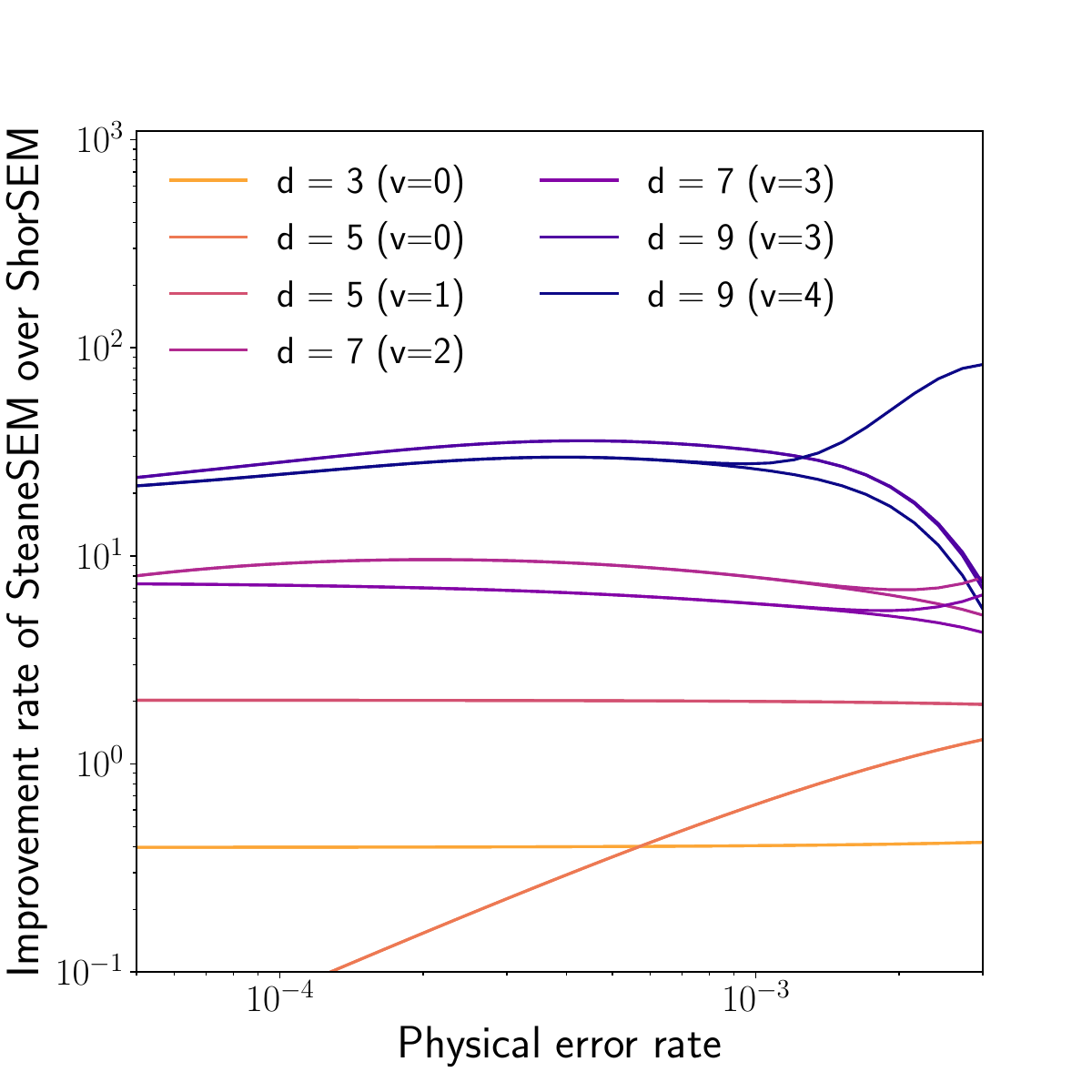}       
    \vspace{0pt}
    \caption{Improvement rate of SteaneSEM over the weak ShorSEM for various distances as a function of the physical error rate. The label v refers to the number of qubits used for the verification of each GHZ state that forms the logical ancillary $|0\rangle_L$ and $|+\rangle_L$ states.  The improvement rate increases monotonically with $d$, from $r \approx 0.4$ for $d=3$ (ShorSEM $2.5$ times better than SteaneSEM), to $r \approx 20$ for $d=9$ (SteaneSEM $20$ times better than ShorSEM).  SteaneSEM advantage is observed for $d=7 ($v$=2)$ and $d=9$, despite the fact that these protocols are not formally FT, while their ShorSEM counterparts are.  Even for $d=5$ without verification (v$=0$), SteaneSEM achieves a comparable performance to ShorSEM for high physical error rates.   For $d=7$ and $d=9$, the divergence between the lower and upper bounds indicate that the probability of occurrence of the high-weight subsets not sampled is considerable.}
    \label{advantage}
    \vspace{0pt}
\end{figure}

In order to further compare both syndrome extraction methods, we define the improvement rate as the ratio of the logical error rate for ShorSEM over the logical error rate of SteaneSEM: 
\begin{equation*}
    r = \frac{p_{L (ShorSEM)}}{p_{L (SteaneSEM)}}
\end{equation*}


Figure \ref{advantage} shows the improvement rate $r$ as function of the physical error rate $p$.  It increases with the code distance from less than $1$ for $d=3$ (a disadvantage) to $\sim 2$ for $d=5$, $\sim 10$ for $d=7$, and $\sim 20$ for $d=9$.  The improvement rate does not depend too strongly on the physical error rate for the $p$ interval considered, as seen in Figure \ref{advantage}.  

The increasing improvement rate of SteaneSEM over ShorSEM has very important practical consequences.  Notice, for example, that for a physical error rate $p = 10^{-4}$, a distance-9 BS code with SteaneSEM would be enough to achieve a logical error rate of $10^{-10}$, which is considered as the largest tolerable error rate to achieve an algorithmic qubit capable of sustaining a deep and useful quantum computation \cite{Beverland2022assessing}.  For ShorSEM, on the other hand, a distance-9 BS code would not be enough to achieve this logical error rate.  It is an open question whether this increasing improvement rate will be sustained for even higher distances ($d>9$).   


\begin{figure}
    \centering
    \includegraphics[width=0.45\textwidth]{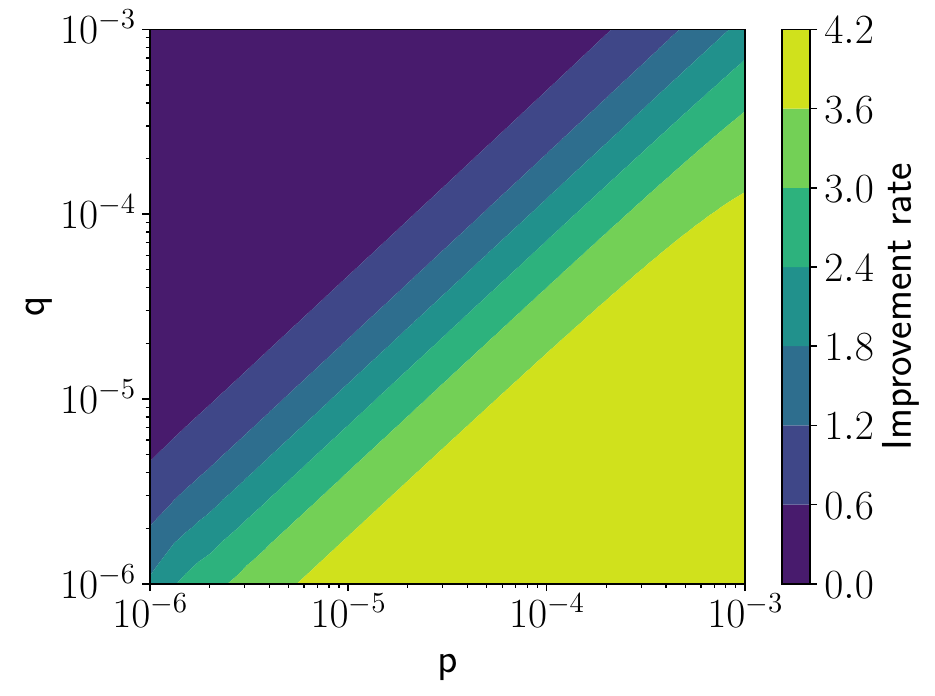}
    \caption{Improvement rate of SteaneSEM over the weak ShorSEM for the distance-5 BS code.  We define two independent noise parameters: $p$, which quantifies the error rate of 1-qubit gates, 2-qubit gates, and state preparations, and $q$, which quantifies the measurement error rate.    The improvement rate is the largest when gate errors dominate over measurement errors ($p \gg q$), which corresponds to the lower right section of the heat map.  In the opposite regime ($p \ll q$), the improvement rate is less than 1, which means that ShorSEM outperforms SteaneSEM.  When $p=q$ (diagonal line), the improvement rate is around 2.}
    \label{fig:advatangeheatmap}
\end{figure}

Motivated by a recent paper \cite{WeileiDataSyndrome}, we now set the measurement error rate ($q$) to be independent from the 1-qubit gate, 2-qubit gate, and state preparation error rate ($p$).  The resulting improvement rate is plotted as a heat map in Figure \ref{fig:advatangeheatmap}.  We find that the improvement rate has the highest values when gate errors dominate over measurement errors ($p \gg q$).  This is expected since the number of entangling gates is considerably lower for SteaneSEM.  However, when measurement errors dominate over gate errors ($p \ll g$), the improvement rate can be less than 1, which means that ShorSEM outperforms SteaneSEM.  This is also expected, since ShorSEM has less measurements than SteaneSEM.  In Appendix \ref{app:NumberGates}, we present the total number of CNOT gates and measurements for the various circuit constructions that we study in this paper.

\subsection{Probability of not passing the GHZ state verification}

\begin{figure}
    \centering
    \includegraphics[width=0.5 \textwidth]{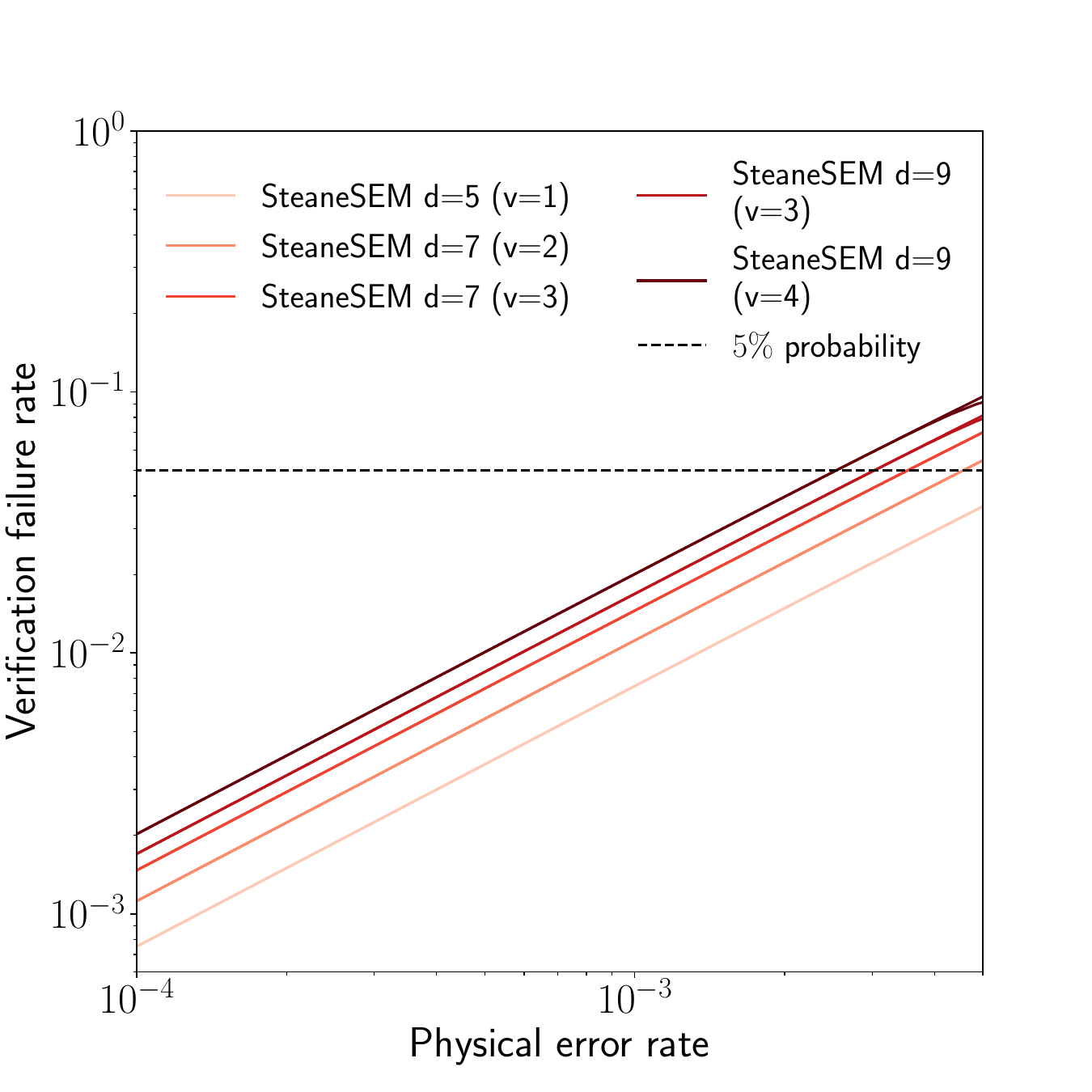}       
    \caption{Probability of discarding the GHZ state because of a failed verification.  The probability grows with increasing physical error rate, but it remains within reasonable limits.  The $5\%$ failure probability is shown as a reference.}
    \label{verificationprob}
\end{figure}

One possible drawback of the current proposal for SteaneSEM on the BS code is that it relies on post-selection.  If the GHZ state does not pass the verification checks, it is discarded and re-prepared, which could potentially result in a considerable time overhead in a system where the measurements and state preparations are slow.  However, a numerical calculation shows that this is not as bad as it might seem.  Figure \ref{verificationprob} shows the probability that a single $d$-qubit GHZ state does not pass the verification.  The probability that the verification is not passed increases with the code distance, but it is very reasonable.  As shown in Figure \ref{verificationprob}, for a physical error rate of $p=10^{-3}$, all the verifications will be passed more than $95\%$ of the times.    

To experimentally implement the SteaneSEM on the BS code, we envision a whole section of the quantum computer fully and exclusively dedicated to preparing and verifying GHZ states.  It would be a continuous-flow factory of verified GHZ states that would then be sent to the section where the QEC step would take place.  This GHZ-state factory would be similar in spirit to the magic state factories proposed for the quantum computing schemes that achieve universality by preparing high-fidelity magic states.  A great advantage of the BS code with SteaneSEM is that it is not necessary to wait for $d$ verified $d$-qubit GHZ states to be simultaneously ready to perform QEC.  Since the GHZ states that form a logical $|0\rangle_L$ or $|+\rangle_L$ are unentangled from each other, coupling these states to the logical data qubit can be done independently, which allows for a fast and not necessarily simultaneous coupling of the GHZ states to the logical data qubit.  For the BS code, this represents a clear advantage of SteaneSEM over KnillSEM \cite{KnillEC}, the other single-shot alternative to the ShorSEM.  The KnillSEM would require the preparation of a logical Bell pair, which would, in turn, require preparing and verifying simultaneously $2d$ $d$-qubit GHZ states.  The total number of gates and measurements would be the same for both schemes, but SteaneSEM would allow for a much more parallelizable execution on the BS code since it does not require all GHZ states to be simultaneously ready to perform QEC.


\section{Conclusions and Outlook}
\label{sec:Conclusions}

In this paper we have studied the performance of the BS code with two different syndrome extraction methods: Shor's and Steane's.  The fact that the logical $|0\rangle_L$ and $|+\rangle_L$ states of the BS code can be expressed as products of GHZ states positions this code as a natural candidate for Steane's syndrome extraction method.  

We have shown that these logical BS states can be prepared in a straightforward manner by verifying its constituent GHZ states and post-selecting them.  By using this post-selection GHZ preparation method we have found that SteaneSEM significantly outperforms ShorSEM on the BS code.  Steane's method results in pseudo-thresholds that are about 1 order of magnitude higher $\left( \sim 10^{-3} \, \textrm{vs.} \, \sim 10^{-4} \right)$.  We have also found that the improvement rate of Steane's method over Shor's increases monotonically with the code distance from Steane's method being disadvantageous for $d=3$ to being about 20 times better for $d=9$.  When we let the measurement error rate be independent from the gates and preparations error rate, we find that Steane's improvement is the greatest in the regime where gate error dominate over measurement errors.  This is consistent with the fact that Steane's method employs considerably less entangling gates than Shor's.

Some of the circuit constructions that we have found to prepare the logical $|0\rangle_L$ and $|+\rangle_L$ states used in Steane's method are not strictly FT, since their effective distance is reduced.  However, we found that for experimentally relevant physical error rates, even the non-FT Steane's circuit constructions outperform their Shor's counterparts, despite the latter ones being formally FT.  Furthermore, for these experimentally relevant physical error rates, Steane's non-FT and FT protocols have essentially the same performance.  From our perspective, this is one of the most important results from this work, since it illustrates that maintaining the formal code distance is not strictly necessary to guarantee the usefulness of a QEC protocol.  It also suggests that leading-order analysis might not be the most appropriate tool when comparing QEC strategies under experimentally relevant physical error rates.

Given that our state preparation methods post-selective, we also calculated the probability that the GHZ verifications are not passed.  We have found that, for all code distances and experimentally relevant physical error rates, the preparations fail less than $5\%$ of the times.

There are several questions that remain to be answered.  First, the circuits used to verify the GHZ states were found by a brute-force exhaustive search of all possible weight-2 Z stabilizers.  Although it is not strictly necessary to find formally FT verification circuits (since, for realistic physical error rates, the performance of FT and non-FT verification circuits is the same), it would be interesting to apply the great body of work on flag qubits \cite{ChaoFlagPRL, ChaoFlagPRX, Chamberland2018flagfaulttolerant} to develop a formal framework to construct GHZ-verification circuits that employ the least number of extra qubits.  It would also be interesting to find circuits that do not rely on post-selection and to explore how to convert these circuits into measurement-free protocols \cite{GotoMF,PerlinMF,MarkusMF}, which would be advantageous for systems with slow and noisy measurements. 

Another important question is determining, for a realistic physical error rate (for instance, $p = 10^{-3}$), what code distance would be necessary to suppress the logical error rate to a sufficiently low, algorithmically useful value ($p_L < 10^{-10}$).  Achieving this would be impossible for the BS code with ShorSEM, since the pseudo-thresholds are below $10^{-3}$ (see Figure \ref{mainplot}).  However, although the BS code in the current setting has no threshold, it might still be possible to achieve sufficiently low logical error rates with SteaneSEM. 

Finally, we also plan to study these protocols in the context of trapped ions and neutral atoms with a realistic modeling of the noise and the possible shuttling operations.  In designing effective and useful FT QEC protocols, it will be crucial to take into account the detailed noise processes and architectural constraints of the specific quantum computing platforms where these protocols will be implemented.







The simulation toolkit and results are available at \cite{GithubRepo}.

\section{Acknowlegments}

The idea behind this paper originated while Mauricio Guti\'errez was visiting the Duke Quantum Center.  Mauricio Guti\'errez would like to thank Kenneth Brown for hosting him and for his great warmth and hospitality.  He would also like to thank Shilin Huang, for the very fruitful discussions that gave rise to this paper, Theerapat Tansuwannont, for his explanations of the adaptive time decoder for ShorSEM, and B\'alint Pat\'o for helpful and detailed comments on the preprint.  The simulations were run on the computer cluster of the Research Center for Materials Science and Engineering (CICIMA) of the University of Costa Rica.  The authors thank Federico Mu\~noz for his help in setting up the simulations on the computer cluster.   This work was partially supported by the Research Office of the University of Costa Rica.  Both authors are grateful to the University of Costa Rica for its support.

\bibliography{references}

\newpage
\appendix
\begin{widetext}
\section{Calculating the Polynomial Expansion Coefficients} \label{app:Coeff}

To calculate the coefficients of the logical error rate polynomial expansion we can expand Equation \ref{eq:pl} to get a function of the physical error rates and then add all the coefficients corresponding to the term we wish to find.
For the cases presented in this paper, the error subsets have only three indices.  If all errors occur with the same probability $p$, $A_{\vec{w}}$ is given by Equation \ref{eq:aw}.

\begin{align}
    A_{\vec{w}} &= \binom{n_1}{w_1} \binom{n_2}{w_2} \binom{n_3}{w_3} p^{w_{tot}}(1-p)^{n_{tot}-w_{tot}} \nonumber \\
    &= \sum_{r=0}^{n_{tot} - w_{tot}} (-1)^{r} \binom{n_1}{w_1} \binom{n_2}{w_2} \binom{n_3}{w_3} \binom{n_{tot}-w_{tot}}{r} p^{w_{tot}+r}
    \label{eq:aw}
\end{align}

where $n_1$ is the number of 1-qubit gates and state preparations, $n_2$ is the number of 2-qubit gates, and $n_3$ is the number of measurements in the circuit; $w_1$ is the number of errors after 1-qubit gates and state preparations, $w_2$ is the number of errors after 2-qubit gates, and $w_3$ is the number of errors after measurements for the particular error subset $A_{\vec{w}}$; $n_{tot} = n_1 + n_2 + n_3$ and $w_{tot} = w_1 + w_2 + w_3$.   Substituting Equation \ref{eq:aw} into Equation \ref{eq:pl} we get Equation \ref{eq:expansio}.

\begin{equation}
    p_{L}^{(lower)} = \sum_{\vec{w} = (0,0,...,0)}^{\vec{w}_{max}} \sum_{r=0}^{n_{tot}-w_{tot}} (-1)^{r} \binom{n_1}{w_1} \binom{n_2}{w_2} \binom{n_3}{w_3} \binom{n_{tot}-w_{tot}}{r} p^{w_{tot}+r} \, p_L^{(\vec{w})}
    \label{eq:expansio}
\end{equation}

Then we can get the coefficient $c_l$ of the term $p^l$ by adding the coefficients of all the terms that satisfy $w_{tot} + r = l$, as shown equation \ref{eq:coeff}

\begin{align}
    c_{l} &= \sum_{\substack{r=0 \\ w_{tot}+r=l}}^{n_{tot} - w_{tot}} (-1)^{r} p_L^{(\vec{w})} \binom{n_1}{w_1} \binom{n_2}{w_2} \binom{n_3}{w_3} \binom{n_{tot}-w_{tot}}{r}
    \label{eq:coeff}
\end{align}

\section{GHZ Preparation Circuits}  \label{app:Circs}

The circuits used to prepare the GHZ states are shown in figures \ref{fig:d5v1}, \ref{d7v2}, \ref{d7v3}, \ref{d9v3} and \ref{d9v4}. The circuit used to perform ShorSEM for a distance-3 BS code is shown in Figure \ref{fig:shorSEMd3}.

\begin{figure}[ht]
    \centering
    \[\Qcircuit @C=0.01em @R=0.01em @! 
    {
    \lstick{\ket{0}} & \gate{H} & \ctrl{1} & \qw & \qw & \qw & \qw & \qw & \qw\\
    \lstick{\ket{0}} & \qw & \targ & \ctrl{1} & \qw & \qw & \ctrl{4} & \qw & \qw\\
    \lstick{\ket{0}} & \qw & \qw & \targ & \ctrl{1} & \qw & \qw & \qw & \qw\\
    \lstick{\ket{0}} & \qw & \qw & \qw & \targ & \ctrl{1} & \qw & \ctrl{2} & \qw\\
    \lstick{\ket{0}} & \qw & \qw & \qw & \qw & \targ & \qw & \qw & \qw\\
    & & & & & \lstick{\ket{0}} & \targ & \targ & \measuretab{M_{z}}\\
    }\]
    \caption{Circuit used to prepare a distance-5 GHZ state with 1 verification.  This circuit is formally FT.}
    \label{fig:d5v1}
\end{figure}

\begin{figure}[h]
    \centering
    \[\Qcircuit @C=0.01em @R=0.01em @! 
    {
    \lstick{\ket{0}} & \gate{H} & \ctrl{1} & \qw & \qw & \qw & \qw & \qw & \qw & \qw & \qw & \qw & \qw \\
    \lstick{\ket{0}} & \qw & \targ & \ctrl{1} & \qw & \qw & \qw & \qw  & \ctrl{6} & \qw & \qw & \qw & \qw\\
    \lstick{\ket{0}} & \qw & \qw & \targ & \ctrl{1} & \qw & \qw & \qw & \qw & \qw & \ctrl{6} & \qw & \qw \\
    \lstick{\ket{0}} & \qw & \qw & \qw & \targ &\ctrl{1} & \qw & \qw & \qw & \qw & \qw & \qw & \qw\\
    \lstick{\ket{0}} & \qw  & \qw & \qw & \qw & \targ & \ctrl{1} & \qw & \qw & \ctrl{3} & \qw & \qw & \qw\\
    \lstick{\ket{0}} & \qw & \qw & \qw & \qw & \qw & \targ & \ctrl{1} & \qw & \qw & \qw & \ctrl{3} & \qw\\
    \lstick{\ket{0}} & \qw & \qw  & \qw & \qw & \qw & \qw & \targ & \qw & \qw & \qw & \qw & \qw\\
    & & & & & & &\lstick{\ket{0}} & \targ & \targ & \qw & \qw & \measuretab{M_{z}}\\
    & & & & & & &\lstick{\ket{0}} & \qw & \qw & \targ & \targ &\measuretab{M_{z}}\\
    }\]
    \caption{Circuit used to prepare a distance-7 GHZ state with 2 verifications.  This circuit is not formally FT.}
    \label{d7v2}
\end{figure}

\begin{figure}[h]
    \centering
    \[\Qcircuit @C=0.01em @R=0.01em @! 
    {
    \lstick{\ket{0}} & \gate{H} & \ctrl{1} & \qw & \qw & \qw & \qw & \qw & \qw & \qw & \qw & \qw & \qw & \qw & \qw\\
    \lstick{\ket{0}} & \qw & \targ & \ctrl{1} & \qw & \qw & \qw & \qw  & \ctrl{6} & \qw & \ctrl{7} & \qw & \qw & \qw & \qw \\
    \lstick{\ket{0}} & \qw & \qw & \targ & \ctrl{1} & \qw & \qw & \qw & \qw & \qw & \qw & \qw & \qw & \qw & \qw \\
    \lstick{\ket{0}} & \qw & \qw & \qw & \targ &\ctrl{1} & \qw & \qw & \qw & \qw & \qw & \qw & \ctrl{6} & \qw & \qw \\
    \lstick{\ket{0}} & \qw  & \qw & \qw & \qw & \targ & \ctrl{1} & \qw & \qw & \ctrl{3} & \qw & \qw & \qw & \qw & \qw \\
    \lstick{\ket{0}} & \qw & \qw & \qw & \qw & \qw & \targ & \ctrl{1} & \qw & \qw & \qw & \ctrl{3} & \qw & \ctrl{4} & \qw \\
    \lstick{\ket{0}} & \qw & \qw  & \qw & \qw & \qw & \qw & \targ & \qw & \qw & \qw & \qw & \qw & \qw & \qw \\
    & & & & & & &\lstick{\ket{0}} & \targ & \targ & \qw & \qw & \qw & \qw & \measuretab{M_{z}}\\
    & & & & & & &\lstick{\ket{0}} & \qw & \qw & \targ & \targ & \qw & \qw &\measuretab{M_{z}}\\ 
    & & & & & & &\lstick{\ket{0}} & \qw & \qw & \qw & \qw & \targ & \targ &\measuretab{M_{z}}\\
    }\]
    \caption{Circuit used to prepare a distance-7 GHZ state with 3 verifications.  This circuit is formally FT.}
    \label{d7v3}
\end{figure}

\begin{figure}
    \centering
    \[\Qcircuit @C=0.01em @R=0.01em @! 
    {
    \lstick{\ket{0}} & \gate{H} & \ctrl{1} & \qw & \qw & \qw & \qw & \qw & \qw & \qw & \qw & \qw & \qw & \qw & \qw & \qw & \qw \\
    \lstick{\ket{0}} & \qw & \targ & \ctrl{1} & \qw & \qw & \qw & \qw & \qw & \qw & \ctrl{8} & \qw & \qw & \qw & \qw & \qw & \qw \\
    \lstick{\ket{0}} & \qw & \qw & \targ & \ctrl{1} & \qw & \qw & \qw & \qw & \qw & \qw & \qw & \ctrl{8} & \qw & \qw & \qw & \qw & \\
    \lstick{\ket{0}} & \qw & \qw & \qw & \targ &\ctrl{1} & \qw & \qw & \qw & \qw & \qw & \qw & \qw & \qw & \qw & \qw & \qw \\
    \lstick{\ket{0}} & \qw  & \qw & \qw & \qw & \targ & \ctrl{1} & \qw & \qw & \qw & \qw & \qw & \qw & \qw & \ctrl{7} & \qw & \qw \\
    \lstick{\ket{0}} & \qw & \qw & \qw & \qw & \qw & \targ & \ctrl{1} & \qw & \qw & \qw & \qw & \qw & \qw & \qw & \qw & \qw & \\
    \lstick{\ket{0}} & \qw & \qw  & \qw & \qw & \qw & \qw & \targ & \ctrl{1} & \qw & \qw & \ctrl{3} & \qw & \qw & \qw & \qw & \qw \\
    \lstick{\ket{0}} & \qw & \qw & \qw  & \qw & \qw & \qw & \qw & \targ & \ctrl{1} & \qw & \qw & \qw & \ctrl{3} & \qw & \ctrl{4} & \qw \\
    \lstick{\ket{0}} & \qw & \qw & \qw & \qw  & \qw & \qw & \qw & \qw & \targ & \qw & \qw & \qw & \qw & \qw & \qw & \qw\\
    & & & & & & & & &\lstick{\ket{0}} & \targ & \targ & \qw & \qw & \qw & \qw & \measuretab{M_{z}}\\
    & & & & & & & & &\lstick{\ket{0}} & \qw & \qw & \targ & \targ & \qw & \qw &\measuretab{M_{z}}\\ 
    & & & & & & & & &\lstick{\ket{0}} & \qw & \qw & \qw & \qw & \targ & \targ &\measuretab{M_{z}}\\
    }\]
    \caption{Circuit used to prepare a distance-9 GHZ state with 3 verifications.  This circuit is not formally FT.}
    \label{d9v3}
\end{figure}

\begin{figure}
    \centering
    \[\Qcircuit @C=0.01em @R=0.01em @! 
    {
    \lstick{\ket{0}} & \gate{H} & \ctrl{1} & \qw & \qw & \qw & \qw & \qw & \qw & \qw & \qw & \qw & \qw & \qw & \qw & \qw & \qw & \qw & \qw \\
    \lstick{\ket{0}} & \qw & \targ & \ctrl{1} & \qw & \qw & \qw & \qw & \qw & \qw & \ctrl{8} & \qw & \qw & \qw & \qw & \qw & \qw & \qw & \qw \\
    \lstick{\ket{0}} & \qw & \qw & \targ & \ctrl{1} & \qw & \qw & \qw & \qw & \qw & \qw & \qw & \ctrl{8} & \qw & \qw & \qw & \qw & \qw & \qw \\
    \lstick{\ket{0}} & \qw & \qw & \qw & \targ &\ctrl{1} & \qw & \qw & \qw & \qw & \qw & \qw & \qw & \qw & \ctrl{8} & \qw & \qw & \qw & \qw \\
    \lstick{\ket{0}} & \qw  & \qw & \qw & \qw & \targ & \ctrl{1} & \qw & \qw & \qw & \qw & \ctrl{5} & \qw & \qw & \qw & \qw & \ctrl{8} & \qw & \qw \\
    \lstick{\ket{0}} & \qw & \qw & \qw & \qw & \qw & \targ & \ctrl{1} & \qw & \qw & \qw & \qw & \qw & \qw & \qw & \ctrl{6} & \qw & \qw & \qw \\
    \lstick{\ket{0}} & \qw & \qw  & \qw & \qw & \qw & \qw & \targ & \ctrl{1} & \qw & \qw & \qw & \qw & \qw & \qw & \qw & \qw & \qw & \qw\\
    \lstick{\ket{0}} & \qw & \qw & \qw  & \qw & \qw & \qw & \qw & \targ & \ctrl{1} & \qw & \qw & \qw & \ctrl{3} & \qw & \qw & \qw & \ctrl{5} & \qw \\
    \lstick{\ket{0}} & \qw & \qw & \qw & \qw  & \qw & \qw & \qw & \qw & \targ & \qw & \qw & \qw & \qw & \qw & \qw & \qw & \qw & \qw \\
    & & & & & & & & &\lstick{\ket{0}} & \targ & \targ & \qw & \qw & \qw & \qw & \qw & \qw & \measuretab{M_{z}}\\
    & & & & & & & & &\lstick{\ket{0}} & \qw & \qw & \targ & \targ & \qw & \qw & \qw & \qw &\measuretab{M_{z}}\\ 
    & & & & & & & & &\lstick{\ket{0}} & \qw & \qw & \qw & \qw & \targ & \targ & \qw & \qw &\measuretab{M_{z}}\\
    & & & & & & & & &\lstick{\ket{0}} & \qw & \qw & \qw & \qw & \qw & \qw & \targ & \targ &\measuretab{M_{z}}\\
    }\]
    \caption{Circuit used to prepare a distance-9 GHZ state with 4 verifications.  This circuit is not formally FT.}
    \label{d9v4}
\end{figure}

\section{Lookup tables}  \label{app:LookupTables}
For the space decoding of the BS code we used lookup tables. For distances 3 and 5 the lookup tables used are shown in Table \ref{ltd3} and Table \ref{ltd5} respectively.  The lookup tables for distances 7 and 9 are analogous.  They correspond to the lookup tables of a repetition code.  For each distance $d$, the X (Z) stabilizers correspond to weight-$2d$ vertical (horizontal) rectangles.

\begin{table}[h]
    \centering
    \subfigure[X stabilizers to correct Z errors.]{
        \begin{tabular}{|l|l|l|}
            \hline
            $S_{X,1}$ & $S_{X,2}$ & Correction \\ \hline
            0     & 0     & $I$        \\ \hline
            0     & 1     & $Z_3$      \\ \hline
            1     & 0     & $Z_1$      \\ \hline
            1     & 1     & $Z_2$      \\ \hline
        \end{tabular}
    }
    \hfill
    \subfigure[Z stabilizers to correct X errors.]{
        \begin{tabular}{|l|l|l|}
            \hline
            $S_{Z,1}$ & $S_{Z,2}$ & Correction \\ \hline
            0     & 0     & $I$        \\ \hline
            0     & 1     & $X_7$      \\ \hline
            1     & 0     & $X_1$      \\ \hline
            1     & 1     & $X_4$      \\ \hline
        \end{tabular}
    }
    \caption{Lookup tables for the distance-3 BS code. $S_{X,1} = X_1X_2X_4X_5X_7X_8$, $S_{X,2} = X_2X_3X_5X_6X_8X_9$, $S_{Z,1} = Z_1Z_2Z_3Z_4Z_5Z_6$, $S_{Z,2} = Z_4Z_5Z_6Z_7Z_8Z_9$.}
    \label{ltd3}
\end{table}

\begin{table*}[h]
    \centering
    \subfigure[X stabilizers to correct Z errors.]{
        \begin{tabular}{|l|l|l|l|l|}
            \hline
            $S_{X,1}$ & $S_{X,2}$ & $S_{X,3}$ & $S_{X,4}$ & Correction \\ \hline
            0         & 0         & 0         & 0         & $I$        \\ \hline
            0         & 0         & 0         & 1         & $Z_5$      \\ \hline
            0         & 0         & 1         & 0         & $Z_4Z_5$   \\ \hline
            0         & 0         & 1         & 1         & $Z_4$      \\ \hline
            0         & 1         & 0         & 0         & $Z_1Z_2$   \\ \hline
            0         & 1         & 0         & 1         & $Z_3Z_4$   \\ \hline
            0         & 1         & 1         & 0         & $Z_3$      \\ \hline
            0         & 1         & 1         & 1         & $Z_3Z_5$   \\ \hline
            1         & 0         & 0         & 0         & $Z_1$      \\ \hline
            1         & 0         & 0         & 1         & $Z_1Z_5$   \\ \hline
            1         & 0         & 1         & 0         & $Z_2Z_3$   \\ \hline
            1         & 0         & 1         & 1         & $Z_1Z_4$   \\ \hline
            1         & 1         & 0         & 0         & $Z_2$      \\ \hline
            1         & 1         & 0         & 1         & $Z_2Z_5$   \\ \hline
            1         & 1         & 1         & 0         & $Z_1Z_3$   \\ \hline
            1         & 1         & 1         & 1         & $Z_2Z_4$   \\ \hline
        \end{tabular}
    }
    \hfill
    \subfigure[Z stabilizers to correct X errors.]{
        \begin{tabular}{|l|l|l|l|l|}
            \hline
            $S_{Z,1}$ & $S_{Z,2}$ & $S_{Z,3}$ & $S_{Z,4}$ & Correction \\ \hline
            0         & 0         & 0         & 0         & $I$        \\ \hline
            0         & 0         & 0         & 1         & $X_{21}$   \\ \hline
            0         & 0         & 1         & 0         & $X_{16}X_{21}$ \\ \hline
            0         & 0         & 1         & 1         & $X_{16}$   \\ \hline
            0         & 1         & 0         & 0         & $X_1X_6$   \\ \hline
            0         & 1         & 0         & 1         & $X_{11}X_{16}$ \\ \hline
            0         & 1         & 1         & 0         & $X_{11}$   \\ \hline
            0         & 1         & 1         & 1         & $X_{11}X_{21}$ \\ \hline
            1         & 0         & 0         & 0         & $X_1$      \\ \hline
            1         & 0         & 0         & 1         & $X_1X_{21}$ \\ \hline
            1         & 0         & 1         & 0         & $X_6X_{11}$ \\ \hline
            1         & 0         & 1         & 1         & $X_1X_{16}$ \\ \hline
            1         & 1         & 0         & 0         & $X_6$      \\ \hline
            1         & 1         & 0         & 1         & $X_6X_{21}$ \\ \hline
            1         & 1         & 1         & 0         & $X_1X_{11}$ \\ \hline
            1         & 1         & 1         & 1         & $X_6X_{16}$ \\ \hline
        \end{tabular}
    }
    \caption{Lookup tables for distance-5 BS code. The stabilizers are analogous to the distance-3 stabilizers: weight-$10$ vertical (horizontal) rectangles for the X(Z) stabilizers.}
    \label{ltd5}
\end{table*}

\section{Number of gates for the various circuit constructions}  \label{app:NumberGates}

To calculate the total number of gates, for SteaneSEM, we assume that the GHZ verifications are done only once.  Since ShorSEM is adaptive, we do not know \textit{a priori} how many rounds of stabilizer measurements will be run.  The number of gates that we report correspond to the interval between the minimal and maximal number of stabilizer rounds, which are given in Table \ref{tab:repsShorSEM}.

\begin{table}[h]
    \centering
    \begin{tabular}{|l|c|c|}
        \hline
        $d$ & weak ShorSEM & strong ShorSEM \\ \hline
        $3$   & $1-2$ & $2-3$ \\ \hline
        $5$   & $2-4$ & $3-5$ \\ \hline
        $7$   & $3-7$ & $4-8$ \\ \hline
        $9$   & $4-10$ & $5-11$ \\ \hline
    \end{tabular}
    \caption{Minimal and maximal number of rounds of stabilizer measurements for the weak and strong FT ShorSEM protocols.  For the weak FT protocol, the lower bound is equal to $t$ while the upper bound is equal to $\lfloor (t+3)^2/4 \rfloor - 2$, where $t = \lfloor (d-1)/2 \rfloor$ and $d$ is the code distance.  For the strong FT protocol, the lower bound is $t+1$ and the upper bound is $\lfloor (t+3)^2/4 \rfloor - 1$.  Reference \cite{InkAdaptive} offers a detailed and rigorous explanation of these bounds.}
    \label{tab:repsShorSEM}
\end{table}

\begin{table}[h]
\centering
\begin{tabular}{|l|ccccc|cc|}
\hline
\multicolumn{1}{|c|}{\multirow{2}{*}{$d$}} & \multicolumn{5}{c|}{Steane}                                                                                     & \multicolumn{2}{c|}{Shor}          \\ \cline{2-8} 
\multicolumn{1}{|c|}{}                   & \multicolumn{1}{c|}{$v=0$} & \multicolumn{1}{c|}{$v=1$} & \multicolumn{1}{c|}{$v=2$} & \multicolumn{1}{c|}{$v=3$} & $v=4$ & \multicolumn{1}{c|}{weak} & strong \\ \hline
3                                        & \multicolumn{1}{c|}{30}  & \multicolumn{1}{c|}{--}  & \multicolumn{1}{c|}{--}  & \multicolumn{1}{c|}{--}  & --  & \multicolumn{1}{c|}{$24-48$}   & $48-72$     \\ \hline
5                                        & \multicolumn{1}{c|}{90}  & \multicolumn{1}{c|}{110} & \multicolumn{1}{c|}{--}  & \multicolumn{1}{c|}{--}  & --  & \multicolumn{1}{c|}{$160-320$}  & $240-400$    \\ \hline
7                                        & \multicolumn{1}{c|}{--}  & \multicolumn{1}{c|}{--}  & \multicolumn{1}{c|}{238} & \multicolumn{1}{c|}{266} & --  & \multicolumn{1}{c|}{$504-1176$} & $672-1344$   \\ \hline
9                                        & \multicolumn{1}{c|}{--}  & \multicolumn{1}{c|}{--}  & \multicolumn{1}{c|}{--}  & \multicolumn{1}{c|}{414} & 450 & \multicolumn{1}{c|}{$1152-2880$} & $1440-3168$   \\ \hline
\end{tabular}
\caption{Total number of CNOT gates for the various protocols that we study in this paper, where $d$ refers to the code distance and $v$ refers to the number of qubits used to verify each GHZ state for SteaneSEM.  For ShorSEM, the ancillary qubits do not need verification.  For SteaneSEM, the numbers presented are given by $2d(d-1 + 2v) + 2d^2$, since there are $d-1$ CNOTs for the preparation of each GHZ state, $2v$ extra CNOTs for its verification, and there $d$ GHZ states for each ancillary logical state.  The extra $2d^2$ CNOTs correspond to the coupling between the data and ancillary qubits.  For ShorSEM, the number of CNOTs is given by $4d(d-1)r$, where $r$ is the number of stabilizer rounds, since each BS code stabilizer is of weight $2d$ and there are $2(d-1)$ of them.  The number of stabilizer rounds is not fixed \textit{a priori} in the adaptive scheme that we have used, so we present the minimal and maximal values.  The values of $r$ are given in Table \ref{tab:repsShorSEM}.}
\end{table}

\begin{table}[h]
\centering
\begin{tabular}{|l|ccccc|cc|}
\hline
\multicolumn{1}{|c|}{\multirow{2}{*}{d}} & \multicolumn{5}{c|}{Steane}                                                                                     & \multicolumn{2}{c|}{Shor}          \\ \cline{2-8} 
\multicolumn{1}{|c|}{}                   & \multicolumn{1}{c|}{$v=0$} & \multicolumn{1}{c|}{$v=1$} & \multicolumn{1}{c|}{$v=2$} & \multicolumn{1}{c|}{$v=3$} & $v=4$ & \multicolumn{1}{c|}{weak} & strong \\ \hline
3                                        & \multicolumn{1}{c|}{18}  & \multicolumn{1}{c|}{--}  & \multicolumn{1}{c|}{--}  & \multicolumn{1}{c|}{--}  & --  & \multicolumn{1}{c|}{$4-8$}    & $8-12$     \\ \hline
5                                        & \multicolumn{1}{c|}{50}  & \multicolumn{1}{c|}{60}  & \multicolumn{1}{c|}{--}  & \multicolumn{1}{c|}{--}  & --  & \multicolumn{1}{c|}{$16-32$}   & $24-40$     \\ \hline
7                                        & \multicolumn{1}{c|}{--}  & \multicolumn{1}{c|}{--}  & \multicolumn{1}{c|}{126} & \multicolumn{1}{c|}{140} & --  & \multicolumn{1}{c|}{$36-84$}   & $48-96$     \\ \hline
9                                        & \multicolumn{1}{c|}{--}  & \multicolumn{1}{c|}{--}  & \multicolumn{1}{c|}{--}  & \multicolumn{1}{c|}{216} & 234 & \multicolumn{1}{c|}{$64-160$}  & $80-176$    \\ \hline
\end{tabular}
\caption{Total number of measurements for the various protocols that we study in this paper, where $d$ refers to the code distance and $v$ refers to the number of qubits used to verify each GHZ state for SteaneSEM.  For ShorSEM, the ancillary qubits do not need verification.  For SteaneSEM, the numbers presented are given by $2dv + 2d^2$, since each one of the $2d$ GHZ states is verified by $v$ qubits, which have to be measured.  The extra $2d^2$ measurements correspond to the final measurements of the logical ancillary states.  For ShorSEM, the number of measurements is given by $2(d-1)r$, where $r$ is the number of stabilizer rounds, since there are $2(d-1)$ stabilizer generators.  The number of stabilizer rounds is not fixed \textit{a priori} in the adaptive scheme that we have used, so we present the minimal and maximal values.  The values of $r$ are given in Table \ref{tab:repsShorSEM}.}
\end{table}

\end{widetext}

\end{document}